\documentclass[paper]{ieice}
\usepackage{subfigure}
\usepackage[pdftex]{graphicx,xcolor}
\usepackage[fleqn]{amsmath}
\usepackage{newtxtext}
\usepackage[varg]{newtxmath}
\usepackage{amssymb}

\usepackage{bm}
\usepackage{algorithm}
\usepackage{subfigure}
\usepackage[noend]{algpseudocode}
\graphicspath{{./fig/}}

\field{}
\title{Extendable NFV-Integrated Control Method Using Reinforcement Learning}
\authorlist{%
\authorentry{Akito Suzuki}{m}{NTT}\MembershipNumber{}
\authorentry{Ryoichi Kawahara}{m}{Toyo}\MembershipNumber{}
\authorentry{Masahiro Kobayashi}{m}{NTT}\MembershipNumber{}
\authorentry{Shigeaki Harada}{m}{NTT}\MembershipNumber{}
\authorentry{Yousuke Takahashi}{m}{NTT}\MembershipNumber{}
\authorentry{Keisuke Ishibashi}{e}{ICU}\MembershipNumber{}
}
\affiliate[NTT]{The authors are with NTT Network Technology Laboratories, NTT Corporation, Musashino-Shi, Tokyo 180-8585, Japan.}
\affiliate[Toyo]{The author is with Faculty of Information Networking for Innovation and Design, Toyo University, Kita-ku, Tokyo 115-0053, Japan.}
\affiliate[ICU]{The author is with Division of Arts and Sciences, College of Liberal Arts, International Christian University, Mitaka-shi, Tokyo 181-0015, Japan.}

\received{2015}{1}{1}
\revised{2015}{1}{1}

\algnewcommand\algorithmicforeach{\textbf{for each}}
\algdef{S}[FOR]{ForEach}[1]{\algorithmicforeach\ #1\ \algorithmicdo}

\begin{document}
\maketitle
\begin{summary}
Network functions virtualization (NFV) enables telecommunications service providers to realize various network services by flexibly combining multiple virtual network functions (VNFs).
To provide such services, an NFV control method should optimally allocate such VNFs into physical networks and servers by taking account of the combination(s) of objective functions and constraints for each metric defined for each VNF type, e.g., VNF placements and routes between the VNFs.
The NFV control method should also be extendable for adding new metrics or changing the combination of metrics.
One approach for NFV control to optimize allocations is to construct an algorithm that simultaneously solves the combined optimization problem. However, this approach is not extendable because the problem needs to be reformulated every time a new metric is added or a combination of metrics is changed.
Another approach involves using an extendable network-control architecture that coordinates multiple control algorithms specified for individual metrics.
However, to the best of our knowledge, no method has been developed that can optimize allocations through this kind of coordination.
In this paper, we propose an extendable NFV-integrated control method by coordinating multiple control algorithms.
We also propose an efficient coordination algorithm based on reinforcement learning. Finally, we evaluate the effectiveness of the proposed method through simulations.
\footnote[0]{This paper is an extended version of our papers presented in~\cite{suzuki2018extendable,suzuki2018society_full,in2016-103_full}. An earlier version of this paper was presented with no review process in~\cite{suzuki2018society_full,in2016-103_full}, which are domestic conferences in Japan.}
\end{summary}

\section{Introduction}\label{sec:intro}
Network functions virtualization (NFV)~\cite{han2015network_full,mijumbi2016management_full} enables telecommunications service providers (TSPs) to provide various network services by flexibly combining multiple virtual network functions (VNFs). These network services in NFV can be provided by combining multiple VNFs (e.g., virtual machines (VMs), intrusion detection systems (IDSs)), each of which is specified for each network service type, between end-to-end hosts such as a content server and a user terminal. To provide such services, a TSP should optimally allocate virtual networks (VNs) consisting of VNFs and routes between the VNFs into underlying physical servers and networks. It has been reported that the inefficient management of network policies (e.g., VNF placements) accounts for $78\%$ of data-center (DC) downtime~\cite{li2013pace_full,cui2016synergistic_full}; therefore, resources must be optimally allocated to provide carrier-grade network services. To avoid inefficient management, the NFV control method should determine the optimal allocations of the VNFs and optimal routes between the VNFs in the network by solving an optimization problem, which is formulated by objective function(s) (e.g., minimizing link-utilization, server-load), and constraint(s) (e.g., upper limit link-bandwidth, server-CPU). However, though a significant amount of research has been conducted~\cite{herrera2016resource_full}, no unified allocation problem has been formulated and solved that takes into account all conditions consisting of combination(s) of objective functions and constraints.

This paper addresses the challenge of developing an NFV-integrated control method, i.e., how to formulate and solve such a unified optimal allocation problem. This problem becomes more difficult due to the increase in the number of control metrics and diversification of objective functions, where the control metric is defined by parameter(s) to characterize the state of a controlled network, e.g., VNF placements and routes between the VNFs. In addition, the NFV-integrated control method should be extendable, i.e., able to handle new control metrics being added or constraints of control metrics being changed. This requirement is crucial because a optimization problem needs to be solved quickly and easily even when a new network service starts or a new constraint needs to be taken into account.

A simple way to achieve adequate extendability is to solve independently pre-specified optimization problems for each control metric. However, if we independently solve each optimization problem taking into account only the problem's constraints, we may not satisfy all the constraints. Hereafter, we call this problem \textbf{control conflict} (see Section~\ref{motivation} for details). To avoid control conflicts, NFV-integrated control methods need to calculate the optimal allocation that satisfies all conditions determined by combination(s) of control metrics.

Previous studies on NFV-integrated control methods can be categorized into two approaches. One is the \textbf{combined} approach~\cite{jiang2012joint_full,yoshida2014morsa_full,jin2015towards_full,cui2016synergistic_full,beck2015coordinated,li2018msv,draxler2018jasper,li2018online}, which builds a specified algorithm that simultaneously solves the combined optimization problem. However, it is not extendable because we need to reconstruct the problem formula every time the combination of control metrics changes. The other is the \textbf{coordinated} approach~\cite{tsagkaris2013survey_full,tsagkaris2015customizable_full,stamou2019autonomic}, which involves using an extendable control architecture that coordinates multiple control algorithms pre-specified for individual control metrics. Though an extendable control architecture has been proposed, this architecture is only a concept and no specific implementation or formulation is described.

In this paper, we propose an \textbf{extendable} NFV-integrated control method based on the \textbf{coordinated} control architecture, which consists of multiple pre-specified control algorithms and a single coordination algorithm between the control algorithms. Our key idea for extendability is modularization that divides a whole system into standardized functional elements and reduces the interdependence among the elements, which is a widely used technique for designing/managing huge complex systems. We first prepare and solve each control algorithm for each control metric and then interactively improve the results using our coordination algorithm. We also propose an efficient coordination algorithm on the basis of reinforcement learning (RL)~\cite{sutton1998reinforcement_full}. The learning makes it possible to learn the strategy for how to find better allocations efficiently from past exploration steps. Our method requires more iterations than the \textbf{combined} approach, but it achieves extendability through the \textbf{coordinated} control architecture.

This paper is structured as follows. Section~\ref{related} describes related work and Section~\ref{motivation} describes our motivation in detail. Section~\ref{method} describes our extendable NFV-integrated control method and an efficient algorithm for our method using RL. Section~\ref{usecase} describes the use cases for our method, the modeling and formulation of the proposed method, and its extendable implementation. Section~\ref{evaluation} evaluates the performance, and Section~\ref{conclusion} concludes the paper.

\section{Related Work}\label{related}
Previous studies on NFV-integrated control methods can be categorized into two approaches: combined and coordinated.

\subsection{Combined approach}
The combined approach~\cite{jiang2012joint_full,yoshida2014morsa_full,jin2015towards_full,cui2016synergistic_full,beck2015coordinated,li2018msv,draxler2018jasper,li2018online} builds a specified algorithm that simultaneously solves the combined optimization problem.

Jiang et al.~\cite{jiang2012joint_full} studied a combined optimization problem of VM placement and routing to minimize traffic costs in an intra-DC. They also proposed an efficient online algorithm in a dynamic environment under changing traffic loads by leveraging and expanding the technique of Markov approximation. However, since the algorithm approximates the combined optimization problem by utilizing the specific problem structure, it is difficult to extend to another use case.

Yoshida et al.~\cite{yoshida2014morsa_full} designed a plug-in architecture to satisfy various requirements related to NFV resources. They also proposed a modified Multi-objective Genetic Algorithm (MOGA) to obtain approximate solutions in reasonable computation time. However, each plug-in should be pre-formulated as a format of objective functions and/or constraints of MOGA. Moreover, they evaluated only one use case and did not mention extendability for adding and/or changing plug-ins.

Jin et al.~\cite{jin2015towards_full} proposed optimization method to minimize the cost of caching, transcoding, and routing functions for cost-efficient video distribution over the future Internet. They developed two algorithms for maximizing total cache hits and minimizing the networking cost. They improve the solution by alternately calculating those two algorithms. However, this method cannot be applied to general use cases due to the control conflict between objective functions (a detailed example is given in Section~\ref{motivation}).

Cui et al.~\cite{cui2016synergistic_full} formulated the Policy-VM Consolidation problem, which can jointly optimize the VM placement as the origin/destination node and VNF placement as the middle node, and the route between VMs via VNF. They also proposed an efficient and synergistic scheme to jointly consolidate VNFs and VM. However, this scheme is heuristic and specified for only one use case, so it is difficult to extend to other use cases.

Herrera et al.~\cite{herrera2016resource_full} survey the research challenges of solving the resource allocation (RA) problem in NFV-based network architectures, and this problem is called the NFV-RA problem. They classify the NFV-RA problems into three stages: VNFs chain composing, VNF forwarding graph embedding, and VNFs scheduling. They mention that these three stages of the NFV-RA problem are related to each other, and a way to coordinate the three stages is a major challenge of the NFV-RA problem. The aim is to optimize the use of resources to improve the performance of the network.

To coordinate NFV-RA problems, various studies have been conducted~\cite{beck2015coordinated,li2018msv,draxler2018jasper,li2018online}. In particular, Li et al.~\cite{li2018online} formulated a typical three-stage coordinated NFV-RA model as a mixed integer programming (MIP) and proposed a heuristic solution called merge-split viterbi (MSV). However, MSV is a specified algorithm that simultaneously solves the combined optimization problem, so it is difficult to extend to other use cases. Other algorithms~\cite{beck2015coordinated,li2018msv,draxler2018jasper} are similar to the case of MSV.

\subsection{Coordinated approach}
The coordinated approach~\cite{tsagkaris2013survey_full,tsagkaris2015customizable_full,stamou2019autonomic} involves using an extendable control architecture that coordinates multiple control algorithms pre-specified for individual control metrics.

Tsagkaris et al.~\cite{tsagkaris2013survey_full,tsagkaris2015customizable_full} and Stamou et al.~\cite{stamou2019autonomic} proposed a hierarchical network control framework for unifying all control when a network contains multiple control metrics. In particular, their proposed architectures~\cite{tsagkaris2015customizable_full,stamou2019autonomic} include a single Autonomic Network Management (ANM) Core and multiple Autonomic Control Loops (ACLs). Each ACL is a control module specialized for one control metric. The ANM Core integrates control of all ACLs and determines whether each ACL appropriately controls each control metric. However, this extendable control architecture is only a concept, and no specific implementation or formulation is described.

\section{Challenges and motivation}\label{motivation}
In this section, we describe the challenges and motivation for extendable NFV-integrated control methods with a concrete use case. We first consider the use case as an example in which we provide a secure-cloud-computing service consisting of routes between VMs via an IDS. In this case, the control metrics are routes, VM placements, and IDS placements, and each control algorithm is pre-formulated (detailed formulation is given in Section~\ref{sec_eq}).

Independently solving each optimization problem leads to the following control conflicts.

\textbf{(1) Conflict between constraints~-~Capacity overload}: Since each control algorithm takes into account only its constraints, all constraints might not be satisfied at the same time. For example, when each problem for each control metric is independently solved at the same time, VMs and IDSs will be allocated on the same server, resulting in server overload.

\textbf{(2) Conflict between objective functions~-~Oscillatory solution}: If each algorithm with a different objective function is conducted independently, the network may become unstable. For example, if the IDS-allocation algorithm to balance server loads and the VM-allocation algorithm to minimize electric power consumption (i.e., the number of powered-on servers) are used independently (e.g., the latter is done after the former repeatedly), most assignment results are repeatedly changed.

The above conflicts can be avoided by sequentially solving each optimization problem for residual resources of each allocated result. However, the obtained results after conducting all the algorithms are not guaranteed to become optimal. This is because, since the result of the previous algorithm is fixed, an inefficient solution may be inevitable. For example, when the results of the physical distance between allocated VMs are long, inefficient routing is inevitable.

The motivation for this paper is to avoid the conflicts and inefficiency described above. In addition, we address the challenge of extendability for changing control metrics that are considered essential in NFV-integrated control. Our goal is to construct an extendable NFV-integrated method, i.e., enabling NFV control metrics to be changed and added without changing each control-algorithm formulation. Though our proposed method cannot solve all the challenges of extendability or cover all use cases, to the best of our knowledge, this is the first paper to tackle this problem, and we expect that more complicated use cases can be solved by enhancing the proposed method.

\section{NFV-integrated control method}\label{method}
We have developed an extendable NFV-integrated control method by coordinating multiple control algorithms. We have also developed an efficient coordination algorithm by using RL to find better solutions with fewer coordinating iterations than the case without RL.

In this section, we first describe the overview and procedure of the proposed method and then give an overview of RL and a formulation of the proposed coordination algorithm.

\begin{figure}[!t]
\centering
\includegraphics[width=0.95\linewidth]{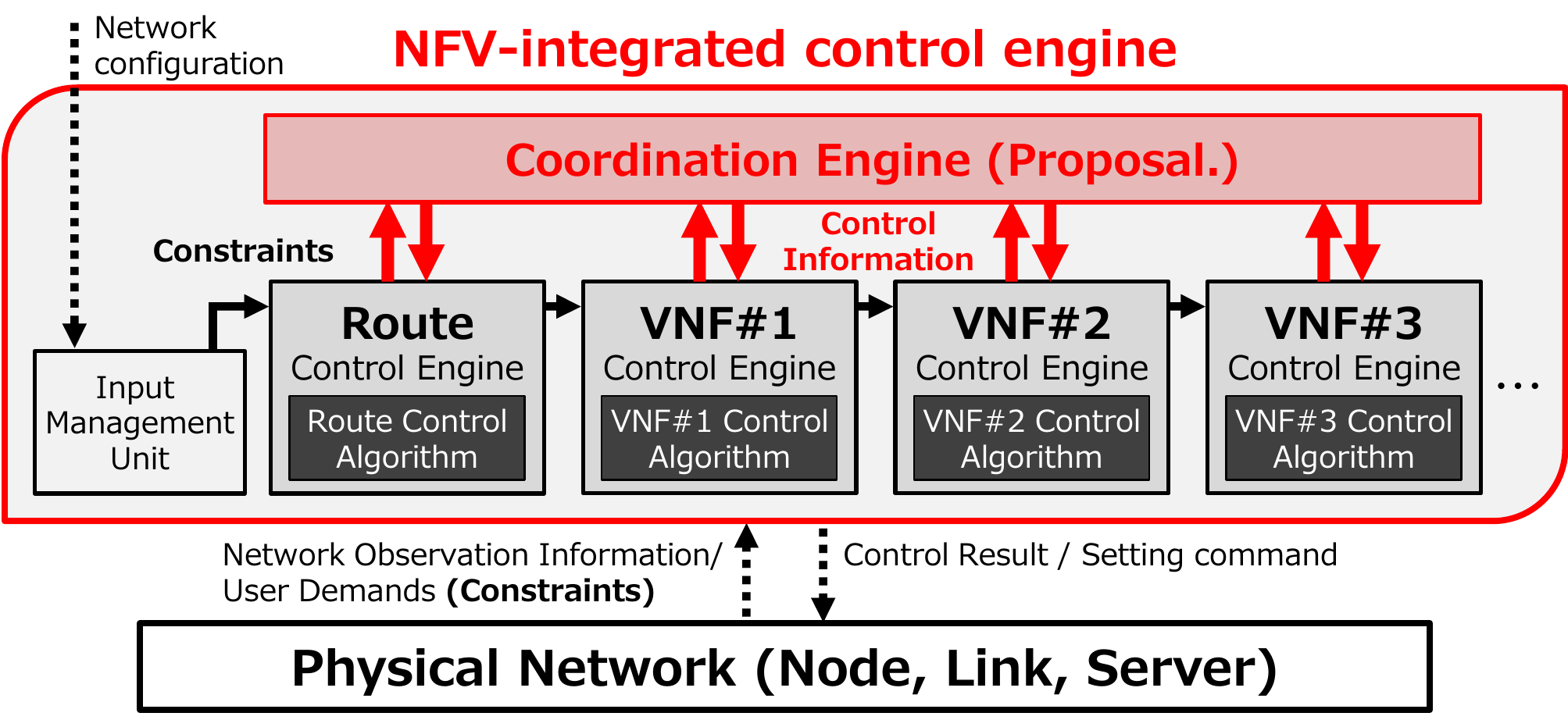}
\caption{Overview of our proposed NFV-integrated control method}
\label{fig1}
\end{figure}

\subsection{Overview of proposed method}
Our method executes hierarchical control consisting of multiple \textbf{control engines} and a single \textbf{coordination engine} (Fig.~\ref{fig1}). A control engine has an algorithm to calculate a solution for each control metric and calculate the evaluation value of the solution quantitatively. The evaluation value of a solution is defined as the objective-function value if the constraints of a control algorithm are satisfied; otherwise, it returns a negative value as a penalty. The objective-function value allows only positive values. Using this negative value, we can determine whether the constraints are satisfied. The coordination engine explores a solution by changing a part of the solution to improve the \textbf{comprehensive evaluation value (CEV)}, which is defined as a unique value determined by all evaluation values of solutions calculated by the individual control engines, e.g., the weighted average of each evaluation value of the solution. The weight of each evaluation value is determined from the importance of each objective function.

We describe the procedure of our proposed method. Each control engine first calculates initial solutions independently, and then our coordination engine recursively explores the solutions to improve the CEV. In the exploration procedure, the coordination engine first changes a part of a solution on the basis of the current CEV, and then the changed solution is sent to each control engine. Next, each control engine calculates the evaluation value on the basis of the changed solution. At this time, some control engines calculate the part of the next solution together as necessary. For example, a route control engine needs to calculate the next route on the basis of the changed VNF placements. The coordination engine calculates the next CEV on the basis of the evaluation value and then returns to the beginning of the procedure. When the exploration is terminated by repeating the above procedure a certain number of times, we regard the highest CEV solution among the past iterations as the final solution. Our method can be extended because we improve each solution on the basis of only the CEV, independently of the control metric type or number of control engines.

\subsection{Overview of coordination algorithm}\label{share}
RL solves the decision problem of what \textbf{action} an ``agent'' should take by observing the current \textbf{state} within a certain ``environment.'' An agent receives a \textbf{reward} from the environment depending on the selected action and then learns a strategy for how to maximize the received reward through a series of selected actions. In an NFV-integrated control environment, an agent's strategy indicates an efficient solution exploration to improve the CEV, and the agent observes the current solution and CEV. We base our proposed algorithm on RL because general-purpose learning is possible by just defining states, actions, and rewards and applying them to general control engines without a specific algorithm.

\begin{figure}[!t]
\centering
\includegraphics[width=0.93\linewidth]{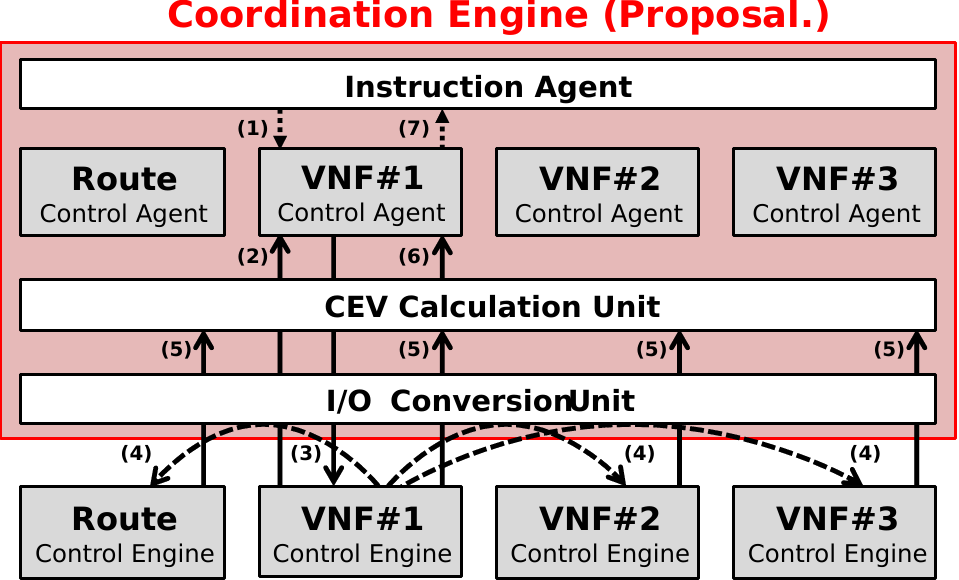}
\caption{Overview of coordination engine based on reinforcement learning. An example when instruction agent selects VNF$\#1$ control agent.}
\label{fig2}
\end{figure} 

Specifically, we use hierarchical multi-agent RL \cite{sun2000self_full}, consisting of a single \textbf{instruction agent} and multiple \textbf{control agents} (Fig.~\ref{fig2}). The instruction agent learns a selection of the control agent, and the control agent learns an efficient solution exploration for the corresponding control engine. The purpose of hierarchical multi-agent RL is to make our method extendable by preparing a specific control agent for each control engine. Since all agents' learning algorithms are common, the learning algorithm is not affected by any change of any control algorithm.

We introduce the input/output (I/O)-conversion unit, which converts the I/O format of each control engine. Because the changed solution of a control engine may affect evaluation values of other control engines, we need to share the changed solution by converting the I/O format. For example, the VNF placements affect the routes between the VNFs, so this unit needs to convert the VNF placements (i.e., an output of a VNF control engine) into traffic demands between servers on which the VNFs are allocated (i.e., input of a route control engine). We assume that the cost of implementing the I/O conversion unit is lower than the cost of rebuilding each algorithm formulation. An example of the I/O-conversion is described in Section~\ref{sec_pro}.

We describe the procedure of our coordination algorithm based on hierarchical multi-agent RL. As shown in Fig.~\ref{fig2}, the instruction agent first selects a control agent on the basis of the RL (1), and then the selected control agent starts exploring solutions and learning a strategy of exploration (2)--(6). In the exploration step, the selected control agent observes the current solution (2), then changes a part of the solution on the basis of the RL, and sends the changed solution (3). After the changed solution is shared through the I/O-conversion unit (4), each control engine calculates its evaluation value. Next, the CEV is calculated from all evaluation values (5), and then the control agent receives the CEV as a reward (6). Then, the instruction agent learns the strategy of selecting a control agent on the basis of the maximum CEV in the exploration (7) and selects the next agent on the basis of the strategy. After repeating the procedure, the best solution is output as the final solution. Finally, the procedure returns to the beginning (1).

\subsection{Formulation of coordination algorithm}
We describe the formulation of our coordination algorithm. Table~\ref{table1} summarizes the definitions of the variables of our coordination algorithm. For agent learning algorithms, we use Q-learning~\cite{sutton1998reinforcement_full}, which learns the relationship of a state, action, and reward to maximize the policy value. Policy value $Q \left( s_{t}, a_{t} \right)$ is defined as the expectation of the sum of rewards obtained in the future when action $a_t$ is selected in state $s_t$.

\begin{table}[t]
\renewcommand{\arraystretch}{1.3}
\caption{Symbol descriptions for coordination algorithm}
\label{table1}
\centering
\begin{tabular}{l||l}
\hline
{\bf Symbol} & {\bf Definition} \\
\hline
\hline
$\rm ia$ & Instruction agent \\
${\bm G} := \left\{ g \right\}$ & Control agent set \\
${\bm E} := \left\{ e \right\}$ & Control engine set \\
$t \in T$ & Exploration step ($T$: Total exploration steps) \\
$t^{g}$ & Number of iterations of control agent ($g \in {\bm G}$) \\
$T^{g}$ & Total exploration steps of control agent ($g \in {\bm G}$) \\
\hline
$s^{\rm agent}_t$ & State of each agent at step $t$ (${\rm agent} \in {\rm ia} \cup {\bm G}$) \\
$a^{\rm agent}_t$ & Action of each agent at step $t$ (${\rm agent} \in {\rm ia} \cup {\bm G}$) \\
$r^{\rm agent}_t$ & Reward of each agent at step $t$ (${\rm agent} \in {\rm ia} \cup {\bm G}$) \\
$Q \left( s^{\rm agent}_t, a^{\rm agent}_t \right)$ & Policy value for state $s^{\rm agent}_t$ and action $a^{\rm agent}_t$ \\
$\alpha, \gamma$ & Hyper-parameter (Default: $\alpha = 0.2$ and $\gamma = 0.9$) \\
\hline
${\bm A}^{e}$ & Solution of control engine $e$ \\
${\bm D_t} := \left\{ d_{ijt} \right\}$ & Traffic demands from node $i$ to node $j$ at step $t$ \\
${\bm V_t} := \left\{ v^e_t \right\}$ & Evaluation values of control engine $e$ at step $t$ \\
${\bm \theta} := \left\{ \theta^e \right\}$ & Coefficients of control engine $e$ \\
\hline
\end{tabular}
\end{table}

\subsubsection{Instruction-agent algorithm}
The instruction agent learns how to select control agents. A state is defined as the selected control agent, action as the selection of the next control agent, and reward as the maximum CEV obtained during this control-agent selection.

The instruction-agent algorithm is shown in Algorithm~\ref{alg1}. Lines $1$--$2$ show the initialization of $Q \left( s^{\rm ia}, a^{\rm ia} \right)$, exploration step $t$, and initial state $s^{\rm ia}_0$. The term ``$\epsilon~{\rm greedy}$'' in line $4$ means the action selected on the basis of the strategy that a random action is selected with probability $\epsilon$; otherwise, an action $a^{\rm ia}_{t}$ that maximizes $Q$ is selected (i.e., ${\rm arg}\max_{a'} Q \left( s^{\rm ia}_t, {a'} \right) $) with probability $1-\epsilon$. It indicates the epsilon-greedy algorithm and is to avoid convergence to a local optimum solution. The term ``${\rm action}$'' in line $5$ shows the action of the instruction agent $a^{\rm ia}_{t}$, which means the switch from the old control agent $\tilde{g}$ to the new control agent $g$, that is, $s^{\rm ia}_{t} = \tilde{g}$ and $s^{\rm ia}_{t+1} = g$. The term ``${\rm agent~learning}~(s^{\rm ia}_{t+1})$'' in line $6$ means control-agent learning (Algorithm~\ref{alg2}). The control agent returns the maximum CEV during exploration and the number of exploration steps. Lines $7$--$8$ show instruction-agent learning, which means that $Q \left( s^{\rm ia}, a^{\rm ia} \right)$ is updated from the relationship of state $s^{\rm ia}$, action $a^{\rm ia}$, and reward $r^{\rm ia}$. $\Delta Q$ is called temporal difference error in RL, which indicates the difference between the current reward and the expected reward. At line $9$, $t^g$ means the number of iterations in Algorithm~\ref{alg2}.

\begin{algorithm}[!t]
\caption{Instruction-agent Learning}
\label{alg1}
\begin{algorithmic}[1]
\State \textbf{initialize:} {$Q \left( s^{\rm ia}, a^{\rm ia} \right) \gets 0$, for all $s^{\rm ia}$ and $a^{\rm ia}$}
\State \textbf{initialize:} {$t \gets 0, s^{\rm ia}_0 \gets {\rm random~choice~from~}{\bm G}$}
\While {$t < T$}
\State {$a^{\rm ia}_t \gets {\rm \epsilon ~greedy} \left( s^{\rm ia}_t \right)$}
\State {$s^{\rm ia}_{t+1} \gets {\rm action} \left( a^{\rm ia}_t \right)$}
\State {$r^{\rm ia}_{t+1}, t^{g} \gets {\rm agent~learning} \left( s^{\rm ia}_{t+1} \right) $}
\State {$\Delta Q \gets r^{\rm ia}_{t+1} + \gamma \max_{a'} Q \left( s^{\rm ia}_{t+1}, {a'} \right) - Q \left( s^{\rm ia}_t, a^{\rm ia}_t \right)$}
\State {$Q \left( s^{\rm ia}_t, a^{\rm ia}_t \right) \gets Q \left( s^{\rm ia}_t, a^{\rm ia}_t \right) + \alpha \Delta Q$}
\State {$t \gets t + t^{g}$}
\EndWhile
\end{algorithmic}
\end{algorithm}

\subsubsection{Control-agent algorithm}
The control agent learns how to efficiently change the solution of the control engine. A state is defined as the solution of the control engine, action as the changing of the control solution of each control engine, and reward as the CEV (examples are given in Section~\ref{sec_pro}).

The control-agent algorithm is shown in Algorithm~\ref{alg2}. The control agent $g$ is selected by the instruction agent, i.e., $g$ corresponds to the current state of instruction agent $s^{\rm ia}$. Lines $1$--$2$ show the initialization of each variable, and ${\bm A}^{e}$ means the initial solution corresponding to the control engine $e$. The term ``${\rm action}$'' in line $5$ means the changing part of the solution of the selected control engine $e$. Then it outputs the result to the selected control engine $e$. In line $6$, the result is shared among other control engines through I/O-conversion unit. Lines $7$--$8$ show the calculation of evaluation values of all control engines on the basis of the changed solution. The term ``${\rm CEV~calculation}$'' in line $9$ means the calculation the CEV as the reward of control agent $r^{g}_{t+1}$. The CEV is basically defined as follows:
\begin{equation}
{\rm CEV} = \sum_{e \in {\bm E}} \theta^e v^e_t,
\end{equation}
where $\theta^e$ and $v^e_t$ are the weighting parameter and evaluation value for the control engine $e$, respectively. The term ``${\rm end~state}$'' in line $12$ means the termination condition of control-agent learning, i.e., the state that does not satisfy one or more constraints. That is, after reaching the solution that does not satisfy at least one constraint, the control agent stops the exploration. In lines $13$ and $15$, the control agent returns the maximum CEV during exploration as a reward for the instruction agent.

\begin{algorithm}[!t]
\caption{Control-agent Learning}
\label{alg2}
\begin{algorithmic}[1]
\State \textbf{initialize:} {$Q \left( s^{g}, a^{g} \right) \gets 0$, for all $s^{g}$ and $a^{g}$}
\State \textbf{initialize:} {$s^{g}_0 \gets {\bm A}^{e}$}
\For {$t = 0{\rm ~to~}T^{g}-1$}
\State {$a^{g}_t \gets {\rm \epsilon ~greedy} \left( s^{g}_t \right)$}
\State {$s^{g}_{t+1} \gets {\rm action} \left( a^{g}_t \right)$}
\State {$\bm D_{t+1} \gets {\rm I/O~conversion} \left( s^{g}_{t+1} \right)$}

\ForEach {$e \in \bm E $}
\State {$v^e_{t+1} \gets {\rm evaluation~by~each~control~engine} \left( \bm D_{t+1} \right)$}
\EndFor

\State {$r^{g}_{t+1}\gets {\rm CEV~calculation} \left( {\bm V_{t+1}}, {\bm \theta} \right)$}
\State {$\Delta Q \gets r^{g}_{t+1} + \gamma \max_{a'} Q \left( s^{g}_{t+1}, {a'} \right) - Q \left( s^{g}_t, a^{g}_t \right)$}
\State {$Q \left( s^{g}_t, a^{g}_t \right) \gets Q \left( s^{g}_t, a^{g}_t \right) + \alpha \Delta Q$}

\If {$s^{g}_{t+1}{\rm ~is~}end~state$}
\State \textbf{return} {$\max_{\tau \in \left\{ 0, 1, \cdots, t \right\}} \left\{ r^{g}_\tau \right\}, t + 1$}
\EndIf

\State {$t \gets t + 1$}
\EndFor
\State \textbf{return} {$\max_{\tau \in \left\{ 0, 1, \cdots, T^{g}-1 \right\}} \left\{ r^{g}_\tau \right\}, T^{g}$}
\end{algorithmic}
\end{algorithm}

\section{Use case of proposed method}\label{usecase}
We consider the use cases where the extendable NFV-integrated control method is required, i.e., where the control metrics and network control conditions are changed and added frequently. First, we classified the general use case of NFV control using a combination of three elements: (1) control metric, (2) control objective, and (3) network model. We also prepared four options as representatives of each element. One option is selected from (1) control metric, one option is selected from (2) control objective, and two options are selected from (3) network model. Finally, we consider 12 use cases excluding 4 invalid combinations from the 16 ($=2^4$) combinations.

In Sections~\ref{sec:taxonomy} and \ref{sec:mod_option}, we first describe the taxonomy of general use cases and the modeling of four options. Then we describe the modeling and formulation of the proposed method and its extendable implementation in Sections~\ref{sec:mod_pro} and \ref{sec:imple}, respectively.

\begin{table*}[!t]
\renewcommand{\arraystretch}{1.25}
\caption{Summary of 12 types of use cases combining 4 options}
\label{table_option}
\centering
\begin{tabular}{ll||c|c|c|c|c|c|c|c|c|c|c|c}
\hline
{\bf Options} & & 1 & 2 & 3 & 4 & 5 & 6 & 7 & 8 & 9 & 10 & 11 & 12 \\ 
\hline
\hline
(1) with IDS                    & &\checkmark&\checkmark&\checkmark&\checkmark&\checkmark&\checkmark&\checkmark&\checkmark& & & & \\
(2) with Reliability            & &\checkmark&\checkmark&\checkmark&\checkmark& & & & &\checkmark&\checkmark& & \\
(3A) with Fixed node            & &\checkmark&\checkmark& & &\checkmark&\checkmark & & &\checkmark& &\checkmark& \\
(3B) IDS isolation or sharing & (\checkmark: isolation) &\checkmark& &\checkmark& &\checkmark& &\checkmark& & -- & -- & -- & -- \\
\hline
\end{tabular}%
\end{table*}

\subsection{Taxonomy of general use case}\label{sec:taxonomy}
Several studies~\cite{herrera2016resource_full,yousafzai2017cloud,son2018taxonomy,pires2015virtual} classify VNF/cloud resource control methods and their use cases. Various use cases are composed of a combination of $3$ elements.

\textbf{(1) Control metric}: The control metric is defined by parameters to characterize the state of a controlled network, e.g., VNF types, VNF model (e.g., CPU, memory, and storage), VNF placements, the combination of the VNFs, the order to go through the VNFs and routes between the VNFs, etc. Specifically, each control metric determines the constraints, e.g., link bandwidth, latency, server capacity, the maximum number of chaining VNFs, etc.

\textbf{(2) Control objective}: Control objectives can be categorized as follows: improvement of resources utilization efficiency (e.g., link and server), network performance (e.g., traffic throughput and latency), quality of service/quality of experience (QoS/QoE), an acceptance rate of service demands, energy efficiency, security and reliability, etc. The number of control objectives also depends on the use case. The use cases in NFV often introduce multiple control objectives. It has been reported that $34\%$ of the previous studies on VM placement used the multi-objective approach~\cite{yousafzai2017cloud}.

\textbf{(3) Network model}: The network model is a specific representation of a controlled network and user demands depending on each use case. The example of network model element is as follows: network topology, traffic transport rule (e.g., route splittable or not), node placement rule (e.g., fixed node placement or not), and resource isolation rule (e.g., with or without network slicing), etc.

We describe two cases with different network models as an example. One example case is when assuming the service function chaining (SFC) in a TSP network. In this case, we generally assume the communication between the client as an origin node and the server as middle nodes or a destination node. Then, we model that the client node is fixed because its location such as a company building using an SFC service is predetermined by the client's location, and the server node can migrate because that function is virtualized as a VM or VNF. In our model, the server resources to be allocated to individual VNFs are separated among users for reasons such as the VNF license fee and security. The other example case is when assuming the data transportation in an inter-DC network. In this case, it is assumed that the functions in origin, middle, and destination nodes can migrate because these functions are virtualized as VMs or VNFs. When a user requests multiple VN demands, or when the DC operator or TSP manages all VN demands, the server resources to be allocated to individual VNFs can be shared among VN demands. The maximum numbers of VNFs and concurrent sessions are practically limited due to the constraints of license and cost. Therefore, it can be modeled that one VNF allocated to near the origin node is selected until the maximum number of concurrent sessions is reached.

\subsection{Modeling of use cases and options}\label{sec:mod_option}
We select $4$ options from the above elements to evaluate our proposed method's extendability and coverage for various use cases. Each option is \textbf{(1) with IDS} (i.e., a representative example of adding a control metric), \textbf{(2) with Reliability} (i.e., a representative example of adding a control objective), \textbf{(3A) with Fixed node} and \textbf{(3B) IDS isolation or sharing} (i.e., representative examples of changing network models). Table~\ref{table_option} shows $12$ use cases combining $4$ options. There are only $12$ use cases because the (3B) IDS isolation or sharing option is effective only under (1) with the IDS option.

We first describe the condition of the simplest case (Case \#$12$). In this case, we assume the use case of computing resource optimization in a single DC as an example. The origin and destination nodes are VMs, i.e., both nodes can migrate. The control metrics are routes and VM placements. Link capacity and server capacity are imposed as constraints. Maximum link utilization efficiency and maximum server utilization efficiency are introduced as control objectives. All routes between the origin and destination are splittable. That is, the traffic between an origin--destination (OD) node pair can be split into multiple routes.

Next, we describe the condition of each option. 
The \textbf{(1) with IDS} option adds IDS placements as control metrics. It is an option passing through an IDS between the origin and destination for all user demands. The same as for VM placements, server capacity is imposed as a constraint and maximizing server utilization efficiency is introduced as a control objective for IDS placements.
The \textbf{(2) with Reliability} option adds maximizing total reliability as a control objective. In this study, reliability is defined by the probability that a packet can go between two points. In other words, it is defined by the one minus failure probability. When the route of each OD is splitting, the reliability of each OD is calculated by multiplying the split ratio and reliability of each route. The formulation of total reliability is described in Section~\ref{sec_eq}. 
The \textbf{(3A) with Fixed node} option decides whether the origin and/or destination node is a fixed node or can migrate. An example of a fixed node is a client node. 
The \textbf{(3B) IDS isolation or sharing} option decides whether IDS resources are shared among VNs or not. When isolating IDSs among VNs, the number of IDSs ($N_{\rm ids}$) that need to be allocated is the same as the number of VNs ($N_{\rm VN}$), that is, $N_{\rm ids} = N_{\rm VN}$. When sharing IDSs among VNs, the $N_{\rm ids}$ is less than the $N_{\rm VN}$, that is, $N_{\rm ids} = M < N_{\rm VN}$.

Some previous studies can be classified into $12$ use cases. The policy and VM consolidation method in cloud DC~\cite{cui2016synergistic_full} are similar to Cases \#$5$--\#$8$. The disaster avoidance control method in a TSP network~\cite{saito2017disaster,honda2019nation} is similar to Cases \#$9$--\#$10$. The joint VM placement and routing control method in DC~\cite{jiang2012joint_full} is similar to Cases \#$11$--\#$12$. However, to the best of our knowledge, no method has been developed that corresponds to Cases \#$1$--\#$4$. In addition, \textbf{no method has been developed that can handle all cases with one extendable algorithm.}

\subsection{Modeling of proposed method}\label{sec:mod_pro}
We describe the modeling and formulation of the proposed method on the basis of Case \#$1$ since it is redundant to explain the modeling of $12$ use cases one by one. We consider the use case in which we provide a secure and reliable cloud-computing service consisting of routes between VMs via an IDS. We describe each formulation of the algorithm and modeling of the proposed method. In this case, the control metrics are routes, VM placements, IDS placements, and reliability. Each control algorithm is pre-formulated. The symbols used in the formulation are defined in Table~\ref{table_ce}.

\begin{table}[!t]
\renewcommand{\arraystretch}{1.3}
\caption{Symbol descriptions for control engines}
\label{table_ce}
\centering
\begin{tabular}{l||l}
\hline
{\bf Symbols} & {\bf Definitions} \\
\hline
\hline
$N_{\rm server}$ & Number of servers \\
${\bm N}, {\bm S}, {\bm L}$ & Node set, server set, link set \\
$P({\bm N}, {\bm L}) = P({\bm S},{\bm L})$ & Physical Network graph \\
${\rm link} \left( i, j \right) \in {\bm L}$ & Link from node $i$ to node $j$ \\
$c_{ij}^{\rm link}$ & Link capacity of ${\rm link} \left( i, j \right)$ \\
$c_i^{\rm server}$ & $i^{\rm th}$ server capacity \\
\hline
$N_{\rm VN}$ & Number of VNs \\
$N_{\rm cli}, N_{\rm vm}, N_{\rm ids}$ & Number of clients, VMs, and IDSs \\
${\bm C}, {\bm V}, {\bm I}$ & Client set, VM set, IDS set \\
$c_i^{\rm ids}$ & $i^{\rm th}$ IDS capacity \\
$w_{i}^{\rm vm}$, $w_{j}^{\rm ids}$ & $i^{\rm th}$ VM size, $j^{\rm th}$ IDS size \\
$t^{\rm VN}_{i}$ & OD Traffic demands for $i^{\rm th}$ VN \\
${\bm \Xi}^{\rm cli} := \left\{ {\xi}^{\rm cli}_{ij} \right\}$ & Client node placement ($i^{\rm th}$ client, $j^{\rm th}$ node) \\
\hline
${\bm T}^{\rm node} := \left\{ t_{pq} \right\}$ & Traffic from node $p$ to node $q$ \\
${\bm T}^{\rm vm} := \left\{ t_{ij}^{\rm vm} \right\}$ & Traffic from VM $i$ to VM $j$ \\
$x_{ij}^{pq}$ & Proportion of passed $t_{pq}$ on ${\rm link} \left( i, j \right)$ \\
${U^{\rm link}_{\rm max}}$ & Maximum link utilization \\
\hline
${\bm \Xi}^{\rm vm} := \left\{ {\xi}^{\rm vm}_{ij} \right\}$ & VM allocation ($i^{\rm th}$ VM, $j^{\rm th}$ server) \\
${\bm \Xi}^{\rm ids} := \left\{ {\xi}^{\rm ids}_{ij} \right\}$ & IDS allocation ($i^{\rm th}$ IDS, $j^{\rm th}$ server) \\
${U^{\rm server}_{\rm max}}$ & Maximum server utilization \\
\hline
$r_{ij}^{\rm link}$ & Link reliability of ${\rm link} \left( i, j \right)$ \\
$r_{i}^{\rm node}$ & $i^{\rm th}$ node reliability \\
$R^{\rm total}$ & total reliability \\
\hline
\end{tabular}%
\end{table}

\subsubsection{Network}
We assume that each physical server is connected to each node, that is $P({\bm N}, {\bm L}) = P({\bm S},{\bm L})$. When each VN request is accepted, the amounts of server and link resources consumed depend on the request size.

We assume that there is a certain number of VN requests $N_{\rm vn}$. A VN request consists of one origin (i.e., client) and one destination (i.e., VM), OD traffic demands, and VM size. Each VM is allocated to a physical server. The VM size indicates the processing capacity of the VM request, such as the requested number of CPU cores. We also assume that each OD traffic demand is routed through an IDS, which is also allocated to a physical server. The IDS size also indicates the processing capacity. Note that if the client and IDS for an OD pair are allocated in the same node, the OD traffic demand between the client and IDS on the network is regarded as $0$. Similarly, if VM and IDS for an OD pair are allocated in the same server, the OD traffic demand between them is regarded as 0.

\subsubsection{Control algorithms}\label{sec_eq}
We introduce four control engines: route, VM, IDS, and reliability (${\bm E} = \left\{ \rm {Route, VM, IDS, Reliability} \right\}$). All control engines have pre-specified control algorithms. The calculation procedure of the initial solution is as follows. After the VM and IDS control algorithms calculate the optimal allocations without taking into account the constraints of other control algorithms, the route control algorithm calculates the end-to-end route between VMs via an IDS. Finally, the reliability control algorithm calculates the reliability on the basis of all end-to-end routes.

We introduce three objective functions: minimization of maximum link utilization for route control, minimization of maximum server utilization for VM and IDS controls, and maximization of total reliability for reliability control. We impose three constraints: link capacity for route control, server capacity for VM control, and server capacity for IDS control.

The route control algorithm is formulated as follows:
\begin{eqnarray}
{\rm min}&:& U_{\rm max}^{\rm link} \label{eq_te1} \\ 
{\rm s.t.}&:& \sum_{j:(i,j) \in L} x_{ij}^{pq} - \sum_{j:(j,i) \in L} x_{ji}^{pq} = 0 \\ 
&& \hspace{2.8cm}(\forall p,q \in N, i \neq p, i \neq q) \nonumber \\ 
&& \sum_{j:(i,j) \in L} x_{ij}^{pq} - \sum_{j:(i,j) \in L} x_{ji}^{pq} = 1 \\
&& \hspace{2.8cm}(\forall p,q \in N, i = p) \nonumber \\ 
&& \sum_{p,q \in N} t_{pq} x_{ij}^{pq} \leq c_{ij}^{\rm link} U_{\rm max}^{\rm link} \nonumber \\ 
&& \hspace{2.8cm}(\forall(i,j) \in L, \forall p,q \in N) \\
&& 0 \leq x_{ij}^{pq} \leq 1\; \hspace{0.925cm}(\forall(i,j) \in L, \forall p,q \in N)\\
&& 0 \leq U_{\rm max}^{\rm link} \leq 1 \label{eq_te2}
\end{eqnarray}

This algorithm calculates a routing variable $x_{ij}^{pq}$ to minimize the link utilization $U_{\rm max}^{\rm link}$ while satisfying the constraints in (3)--(7), where $x_{ij}^{pq}$ shows the proportion of passing OD traffic demands $t_{pq}$ on the link $\left( i, j \right)$. Equations (3)--(4) show the traffic flow conservation law. Equation (5) shows the constraint of link capacity. Equations (6)--(7) show the range of variables.

The VM control algorithm is formulated as follows:
\begin{eqnarray}
{\rm min}&:& U_{\rm max}^{\rm server} \label{eq_vm1}\\
{\rm s.t.}&:& \sum_{s_k \in S} {\xi}_{ik}^{\rm vm} = 1\; \hspace{2.55cm}(\forall v_i \in V) \\
&& \sum_{v_i \in V} w_{i}^{\rm vm} {\xi}_{ik}^{\rm vm} \leq c_k^{\rm server} U_{\rm max}^{\rm server}\; \hspace{0.5cm}(\forall s_k \in S) \\
&& {\xi}_{ik}^{\rm vm} \in \left[ 0,1 \right]\\
&& 0 \leq U_{\rm max}^{\rm server} \leq 1 \label{eq_vm2}
\end{eqnarray}

This algorithm calculates an VM allocation variable ${\xi}^{\rm vm}_{ik}$ to minimize the server utilization $U_{\rm max}^{\rm server}$ while satisfying the constraints in (8)--(12), where ${\xi}^{\rm vm}_{ik}$ shows the VM solution in which ${\xi}^{\rm vm}_{ik}$ is $1$ if $i^{\rm th}$ VM is assigned to the $k^{\rm th}$ server; otherwise, $0$. Equation (9) shows the VM conservation law. In other words, it shows that each VM must be allocated to any server. Equation (10) shows the constraint of server capacity. Equations (11)--(12) show the range of variables.

The formulation of the IDS control algorithm replaces $w_{i}^{\rm vm}$ and ${\xi}_{ik}^{\rm vm}$ with $w_{j}^{\rm ids}$ and ${\xi}_{jk}^{\rm ids}$ for that of the VM control algorithm. Similarly, IDS allocation ${\xi}^{\rm ids}_{jk}$ indicates the IDS solution in which ${\xi}^{\rm ids}_{jk}$ is $1$ if $j^{\rm th}$ IDS is assigned to the $k^{\rm th}$ server; otherwise, $0$.


In this study, reliability is defined by the probability that a packet can go between two points. Especially, node reliability is defined by the packet reachable probability from node ingress to node egress. In other words, it is defined by the one minus node failure probability. The link reliability is also similar. The reliability between ODs is defined as the product of each reliability going through each node and each link between ODs.

The reliability control algorithm is formulated as follows:
\begin{eqnarray}
{\rm min}&:& R^{\rm total} \\
&& R^{\rm total} = \frac{1}{\sum_{k=1}^{N_{\rm VN}} t^{\rm VN}_k}\sum_{k=1}^{N_{\rm VN}} {t^{\rm VN}_k R^{\rm VN}_k} \label{r_total} \\
&& R^{\rm VN}_k = \sum_{p \in {\rm path}(k)} r_p \left( \prod_{i \in N_p} r_{i}^{\rm node} \prod_{(i, j) \in L_p} r_{ij}^{\rm link} \right) \label{r_vn}
\end{eqnarray}

As shown in (\ref{r_total}), total reliability $R^{\rm total}$ is defined as the weighted average of each VN reliability $R^{\rm VN}_k$, where the weight is determined by the traffic demand to each VN, i.e., $\sum_{k=1}^{N_{\rm VN}} t^{\rm VN}_k$. Here, $R^{\rm VN}_k$ is calculated as shown in (\ref{r_vn}). We explain this formula. $r_{i}^{\rm node}$ and $r_{ij}^{\rm link}$ indicate the reliability of node $i$ and the link between nodes $i$ and $j$. If each VN has multiple paths, each VN's reliability is calculated using the traffic splitting ratio. For each path $p \in {\rm path} (k)$ of the $k^{\rm th}$ VN, the traffic splitting ratio $r_p$, the set of nodes that the path $p$ passes through  $N_p$, and the set of links that the path $p$ passes through  $L_p$ are defined. The above three parameters are calculated by each VN allocation result, each traffic demand ${\bm T}^{\rm node}$, and the route control engine's solution $x_{ij}^{pq}$.

Note that we selected these control algorithms as examples, which are commonly used in previous studies. In our proposed method, each control algorithm is saved as a model file and can be changed by only changing the model file.

\subsubsection{Coordination algorithm}\label{sec_pro}
We introduce VM and IDS control agents as the control agents (${\bm G} = \left\{ \rm {VM, IDS} \right\}$). The route control agent is not introduced here because any routes between the VNFs are not changed unless the VNF placements change. The route control engine is used only to calculate the part of the CEV by solving the route control algorithm in each step. Similarly, the reliability control agent is not introduced.

The state of a control agent defines the VM or IDS allocation, that is, $s^{\rm VM} = {\bm \Xi}^{\rm vm}$ or $s^{\rm IDS} = {\bm \Xi}^{\rm ids}$. The action of the control agent defines one VM or IDS migration. The VM or IDS to migrate is selected from the most used server, and the destination server is selected on the basis of its agent strategy. Note that, in this action, only one migration is executed, and the VNF control algorithm described using (\ref{eq_vm1})--(\ref{eq_vm2}) is not solved. The reward of the control agent defines the CEV if all constraints are satisfied; otherwise, the penalty is $-100$. The CEV is defined as follows:
\begin{equation}
r^{g}_t = \theta^{\rm link} \left( 1-U^{\rm link}_{\rm max} \right) + \theta^{\rm server} \left( 1-\tilde{U}^{\rm server}_{\rm max} \right) + \theta^{\rm r} R^{\rm total}_t \label{eq_sogo},
\end{equation}
where $\theta^{\rm link}$ and $\theta^{\rm server}$, and $\theta^{\rm r}$ are weighting parameters indicating the importance of each control-objective function. The term $\tilde{U}^{\rm server}_{\rm max}$ is the maximum server utilization after aggregating VM and IDS allocations. The I/O-conversion unit calculates VN allocation results, which are the set of the origin node (or clients placement), middle node (allocated IDS placement), destination server (allocated VM placement), and ${\bm T}^{\rm node}$, on the basis of ${\bm \Xi}^{\rm vm}$, ${\bm \Xi}^{\rm ids}$, ${\bm \Xi}^{\rm cli}$, and $t^{\rm VN}_{i}$.

\subsection{Implementation difference between options}\label{sec:imple}
We describe the implementation differences with and without each option shown in Table~\ref{table_option}. We describe the required additional implementation in comparison with the situation where the implementation of proposed methods for Case \#$12$ is completed. We also indicate that the above implementation can easily be completed.

\textbf{(1) with IDS}: When the option is added, we need to introduce an IDS agent and IDS control engine. The RL algorithm and its modeling of state, action, and reward are the same for the VM agent and IDS agent, so no additional implementation of the Python code is required for adding the IDS agent. Similarly, no additional implementation of the code is required for adding IDS control engine because the objective function and constraints of the IDS engine are the same as those for the VM control engine in this use case. Even if the formulation of IDS control engine is changed, the formulation of that engine is modularized as a file that describes optimization problem formulations, so the engine can be reformulated by changing a few lines of that file.

In I/O-conversion unit calculation, the format of VN allocation results is changed to that of adding the middle server node information. In the CEV calculation, the term $U^{\rm server}_{\rm max}$ of the maximum server utilization is changed to $\tilde{U}^{\rm server}_{\rm max}$, which is that of the value after aggregating VM and IDS allocations. In our implementation, the above change can be developed with a change of about $10$ lines of Python code.

\textbf{(2) with Reliability}: When the option adds, we need to introduce Reliability control algorithms. Though the implementation of the engine is newly required, it is extendable because the existing code does not change. In addition, the formula to calculate CEV slightly needs to be changed from (\ref{eq_sogo2}) to (\ref{eq_sogo}).
\begin{equation}
r^{g}_t = \theta^{\rm link} \left( 1-U^{\rm link}_{\rm max} \right) + \theta^{\rm server} \left( 1-U^{\rm server}_{\rm max} \right) \label{eq_sogo2}
\end{equation}

\textbf{(3A) with Fixed node}: When the option changes, we need to slightly modify the I/O-conversion unit implementation. In the I/O-conversion unit calculation, origin node placement returns from ${\bm \Xi}^{\rm cli}$ if with the Fixed node option, otherwise, it returns ${\bm \Xi}^{\rm vm}$.

\textbf{(3B) IDS isolation or sharing}: When the option changes, we need to modify the I/O-conversion unit implementation. When the IDS isolation condition is used, the I/O-conversion unit returns the specific IDS for each VN demand, that is, the $i^{\rm th}$ IDS is exclusively allocated to the $i^{\rm th}$ VN. If IDS sharing condition, the I/O-conversion unit returns the best IDS selected from IDS set $\bm I$. The best IDS is defined as the IDS that satisfies two conditions: the length of total OD route via IDS is the shortest, and concurrent sessions are less than the IDS capacity $c^{\rm ids}_i$. In our implementation, the above change can be developed with a change of about $10$ lines of Python code.

\section{Evaluation}\label{evaluation}
We evaluated the effectiveness of the proposed algorithm through simulations in terms of solution-exploration speed, difference from the optimal solution, scalability, and extendability. We use the use cases in Section~\ref{usecase} to evaluate the extendable NFV-integrated control method.

We first evaluate the solution-exploration speed to assess whether our method can find the solution with improved CEV within the practical iterations because our method generally seems to need more iterations than the combined approach developed for a specific problem. In addition, we assess whether RL can find better solution efficiently. Then, we also investigate the difference from the optimal solution. After that, we evaluate the scalability of our method to estimate the practical range of $N_{\rm VN}$ where the CEV can be improved within the practical computational time. In addition, we also discuss the extendability of our proposed method. Since the extendability of our method is difficult to evaluate quantitatively, we show that our method makes it possible to solve all use cases. Since it is redundant to discuss all 12 results, we focused on the 6 use cases for which the effects of changing each option need to be discussed (Cases \#$1$, \#$3$, \#$4$, \#$5$, \#$8$, \#$12$). Then, we investigate the differences between the proposed coordinated method and the previous combined method and also discuss how easy/difficult to build and solve the problem. Finally, we discuss the applicability of the proposed method.

\begin{figure}[!t]
\centering
\includegraphics[width=0.95\linewidth]{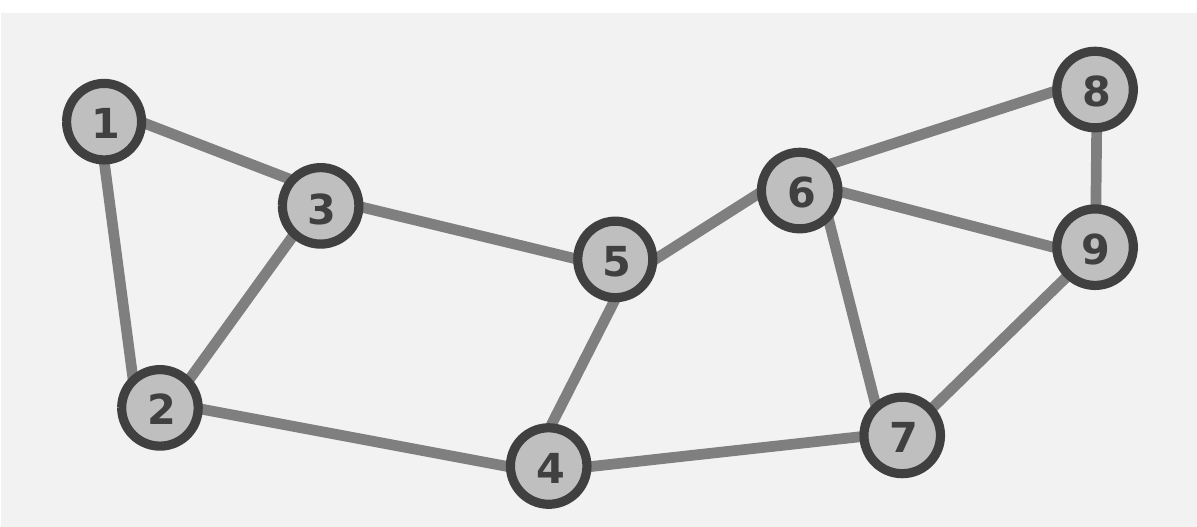}
\caption{
Internet$2$ topology.}
\label{fig3}
\end{figure}

\subsection{Evaluation conditions}\label{sec:eva_conditions}
For the physical network conditions, we used the topology of Internet$2$~\cite{internet2}, which consists of $9$ nodes (Fig.~\ref{fig3}). In particular, we assume a local disaster near node $6$ and set $r_{6}^{\rm node} = 0.9$ and $r_{67}^{\rm link} = r_{69}^{\rm link} = 0.5$. Other $r_{ij}^{\rm link}$ and $r_{i}^{\rm node}$ are set to $1.0$. For the VN demand conditions, the location of each client node is randomly generated. Moreover, the OD traffic demand $t^{\rm VN}_{i}$ is randomly generated within the range of $0$--$1.0$~Gbps so as to arrange the average as $0.5$~Gbps. Each VM size is randomly given an integer value, and each IDS size is fixed to an integer value. The average server utilization is set to $80$\% by changing each server capacity proportionally to the $N^{\rm VN}$ and slightly adjusting the VM size. For the agent conditions, we set the total exploration steps of instruction and control agents to $T = 5000$, and $T^{g} = 20$.

In above conditions, we varied $N_{\rm VN}$ from $20$ to $2000$ and varied use-cases from $1$ to $12$. Some parameters increase proportionally as shown in Table~\ref{table_scale} as the number of VNs are increased from $20$ to $2000$. Some parameters also set depending on the selected use-cases as shown in Table~\ref{table_ext}.
In addition, we excluded the cases in each of which an initial solution does not satisfy all constraints in all evaluations. This is because starting from an unsatisfied initial solution would drastically decrease the performance of the solution. The way to find a feasible initial solution is discussed in Section~\ref{sec:applicability}, which is for future study.

\subsection{Evaluation results}
We implemented our coordination algorithm and physical network simulator using Python from scratch and each pre-specified control algorithm using the GNU Linear Programming Kit (GLPK)~\cite{glpk} to calculate initial solutions.

\begin{table*}[!t]
\renewcommand{\arraystretch}{1.3}
\caption{Scale parameters for Case \#$1$}
\label{table_scale}
\centering
\begin{tabular}{ll||c|c|c|c|c|c|c|c}
\hline
{\bf Definitions} & & 20 & 50 & 200 & 400 & 800 & 1000 & 1500 & 2000 \\
\hline
\hline
Number of VNs & $N_{\rm VN}$                & 20 & 50 & 200 & 400 & 800 & 1000 & 1500 & 2000 \\
Number of VMs & $N_{\rm vm}$                & 20 & 50 & 200 & 400 & 800 & 1000 & 1500 & 2000 \\
Number of IDSs & $N_{\rm ids}$              & 20 & 50 & 200 & 400 & 800 & 1000 & 1500 & 2000 \\
Number of clients & $N_{\rm cli}$           & 20 & 50 & 200 & 400 & 800 & 1000 & 1500 & 2000 \\
$i^{\rm th}$ server capacity & $c_i^{\rm server}$ & 12--14 & 30--36 & 120--144 & 240--288 & 480--576 & 600--720 & 900--1080 & 1200--1400 \\
Link capacity of ${\rm link} \left( i, j \right)$ & $c_{ij}^{\rm link}$ & 3 & 7.5 & 30 & 60 & 120 & 150 & 225 & 300 \\
\hline
\end{tabular}
\end{table*}

\begin{figure*}[!t]
\centering
\subfigure[$N_{\rm VN} = 20$]{
    \includegraphics[width=0.235\linewidth]{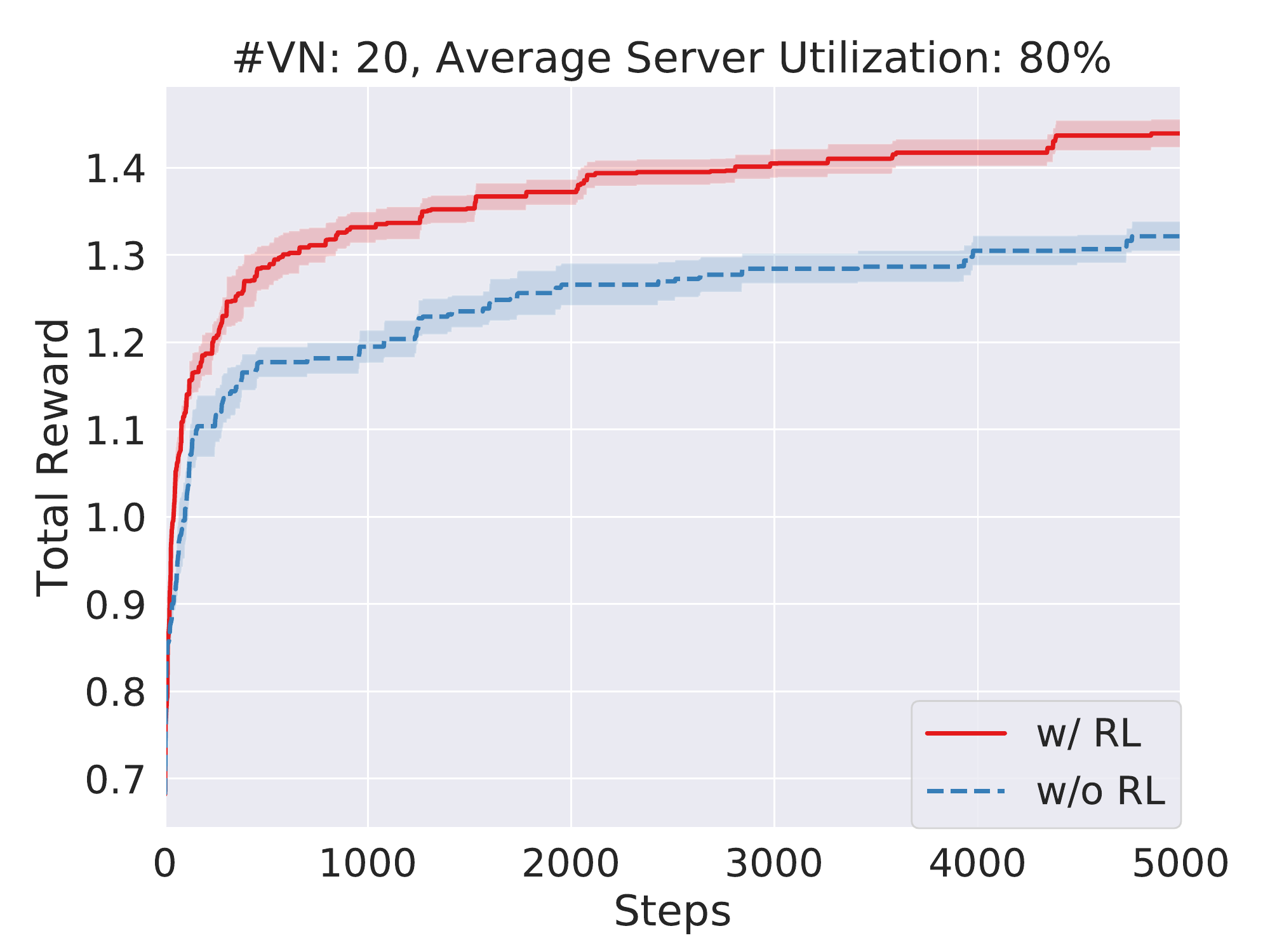}
    \label{fig:20tr}
}
\subfigure[$N_{\rm VN} = 50$]{
    \includegraphics[width=0.235\linewidth]{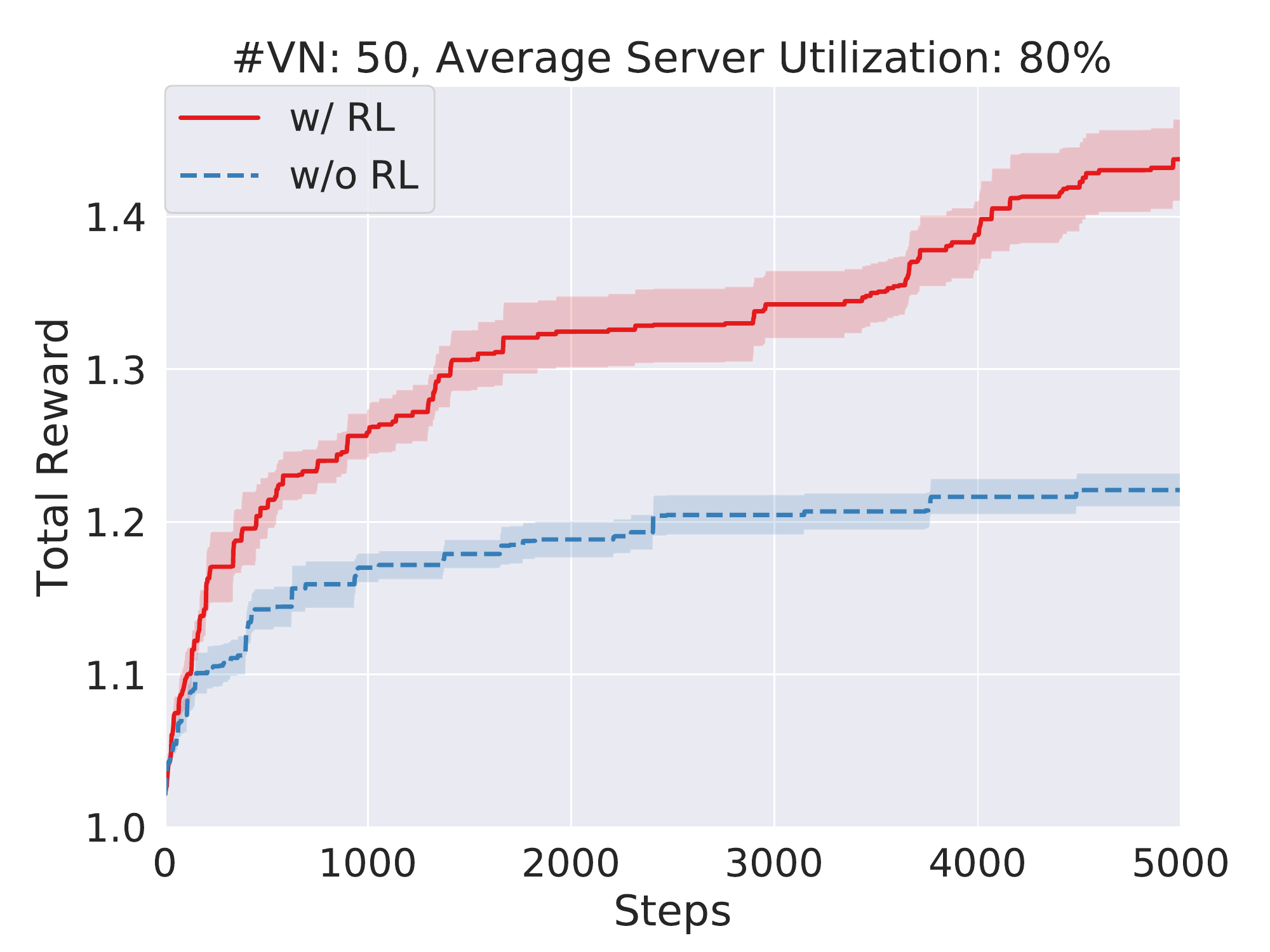}
    \label{fig:50tr}
}
\subfigure[$N_{\rm VN} = 200$]{
    \includegraphics[width=0.235\linewidth]{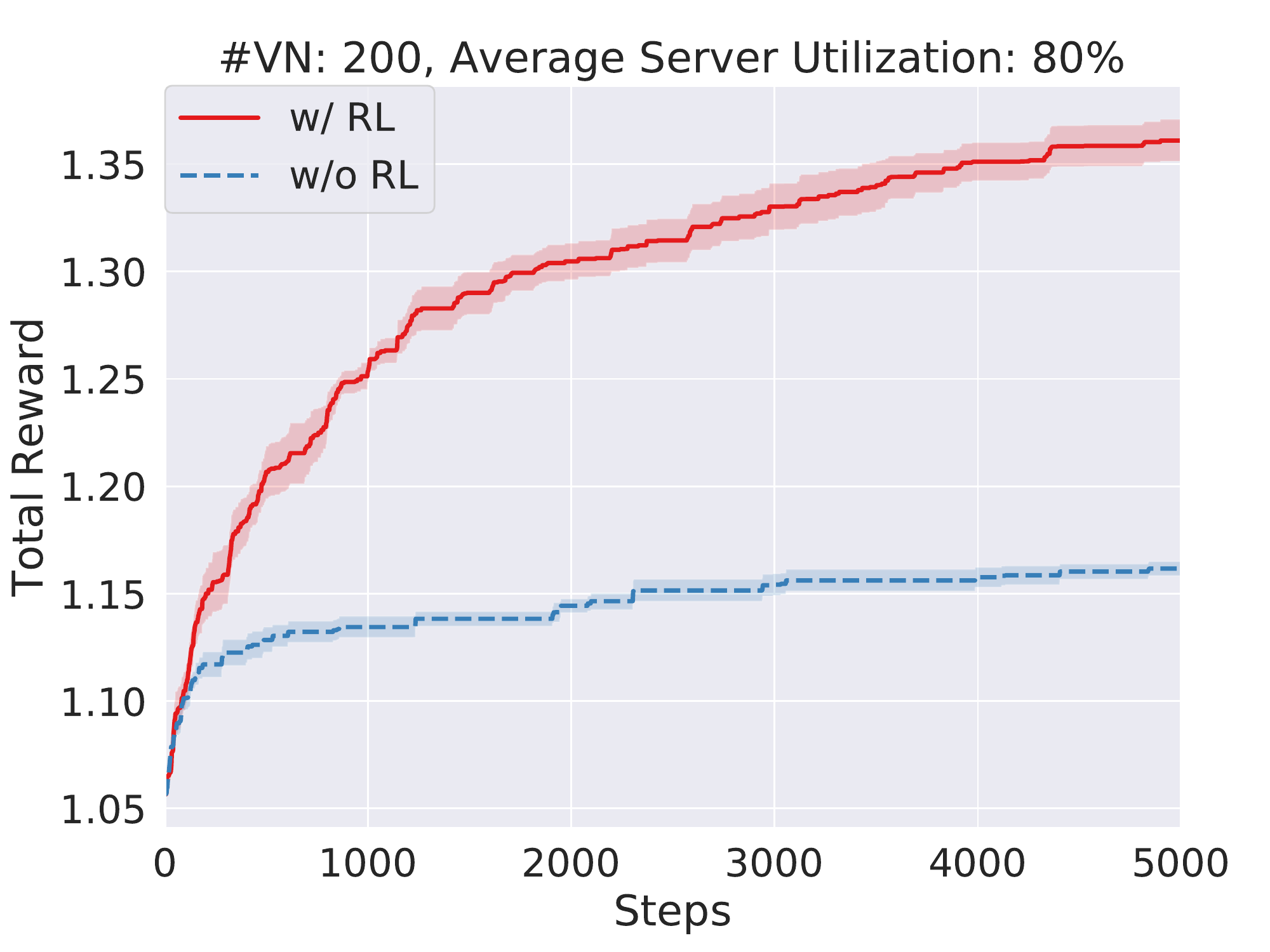}
    \label{fig:200tr}
}
\subfigure[$N_{\rm VN} = 400$]{
    \includegraphics[width=0.235\linewidth]{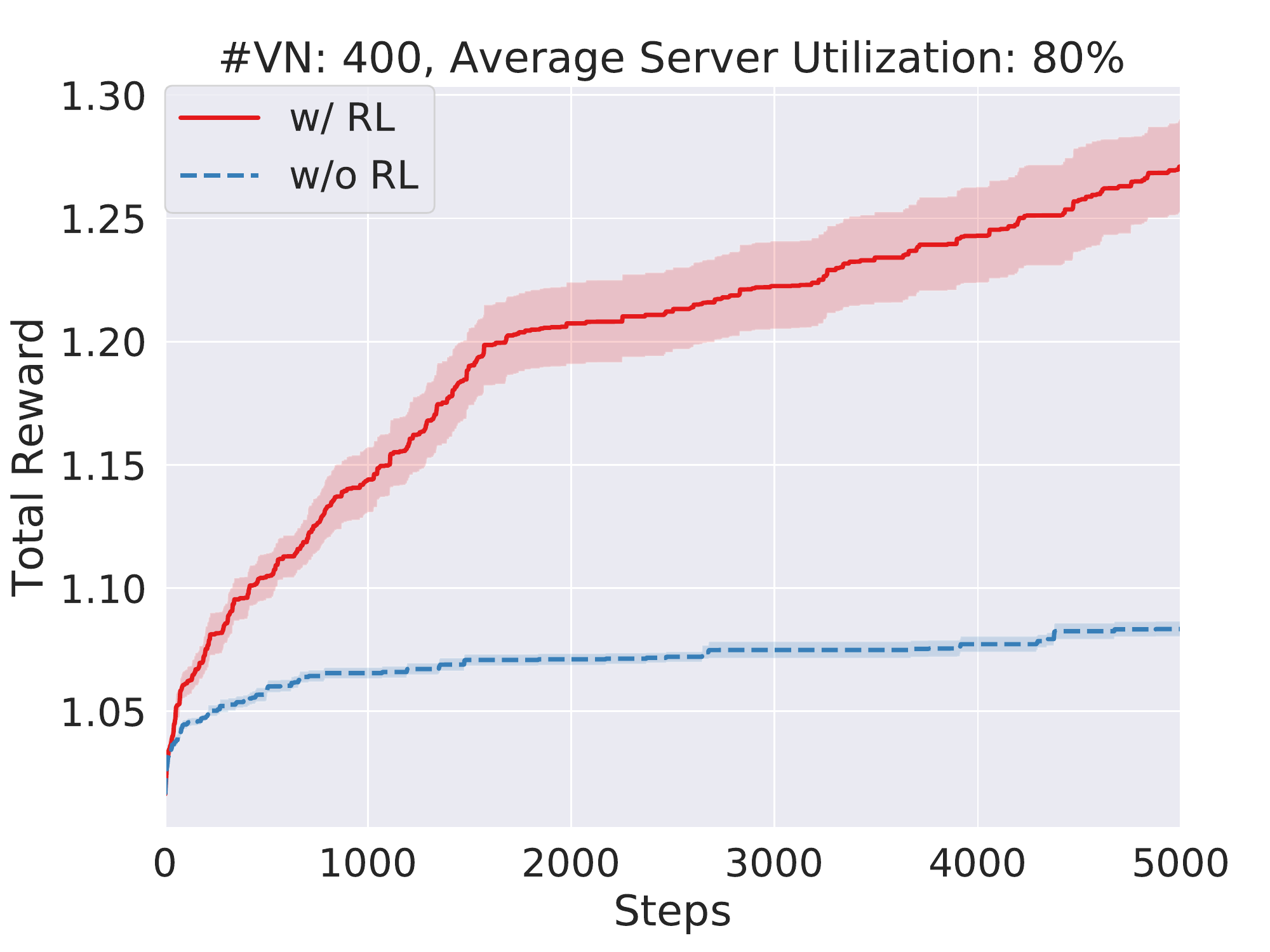}
    \label{fig:400tr}
}
\subfigure[$N_{\rm VN} = 800$]{
    \includegraphics[width=0.235\linewidth]{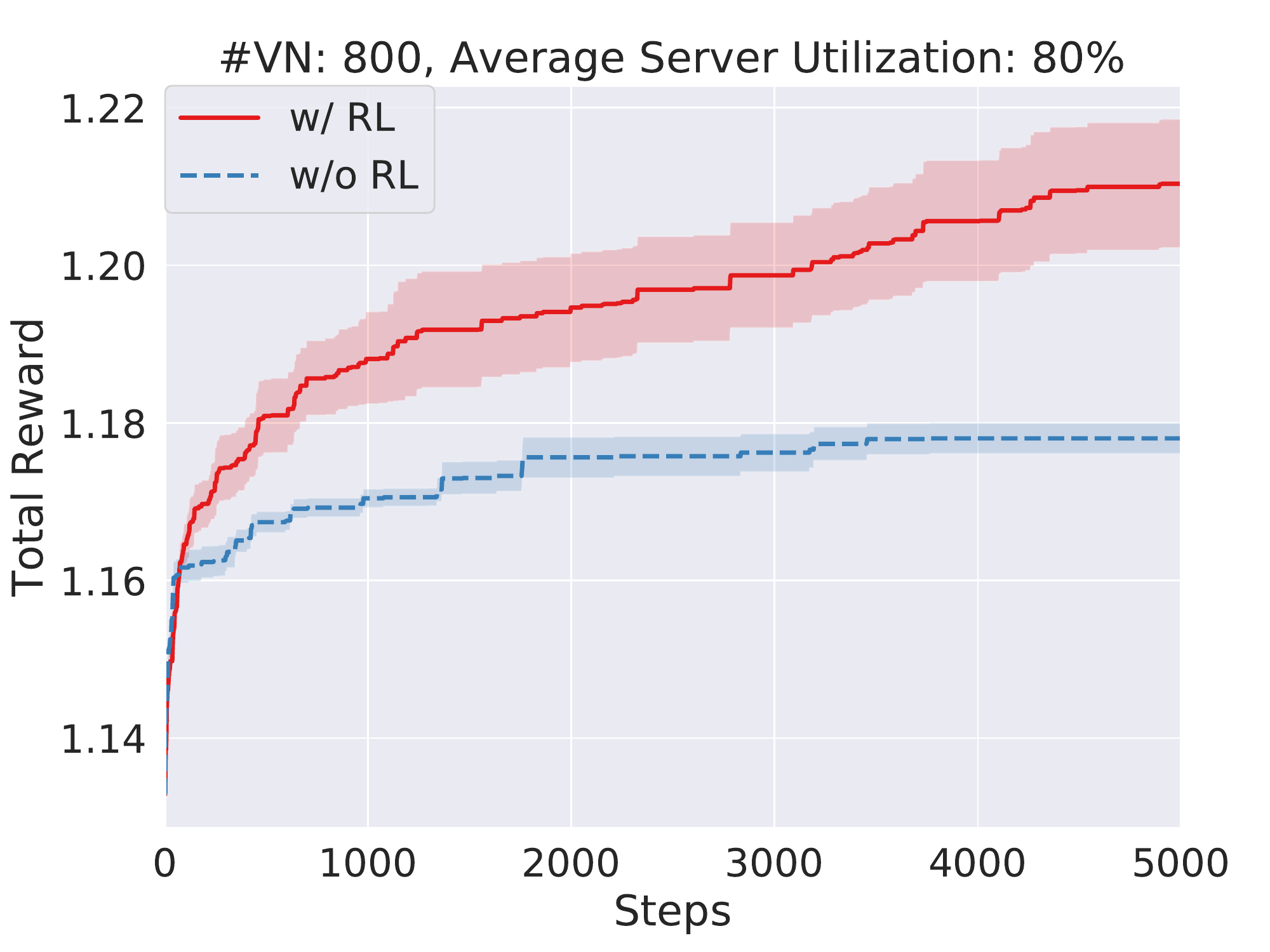}
    \label{fig:800tr}
}
\subfigure[$N_{\rm VN} = 1000$]{
    \includegraphics[width=0.235\linewidth]{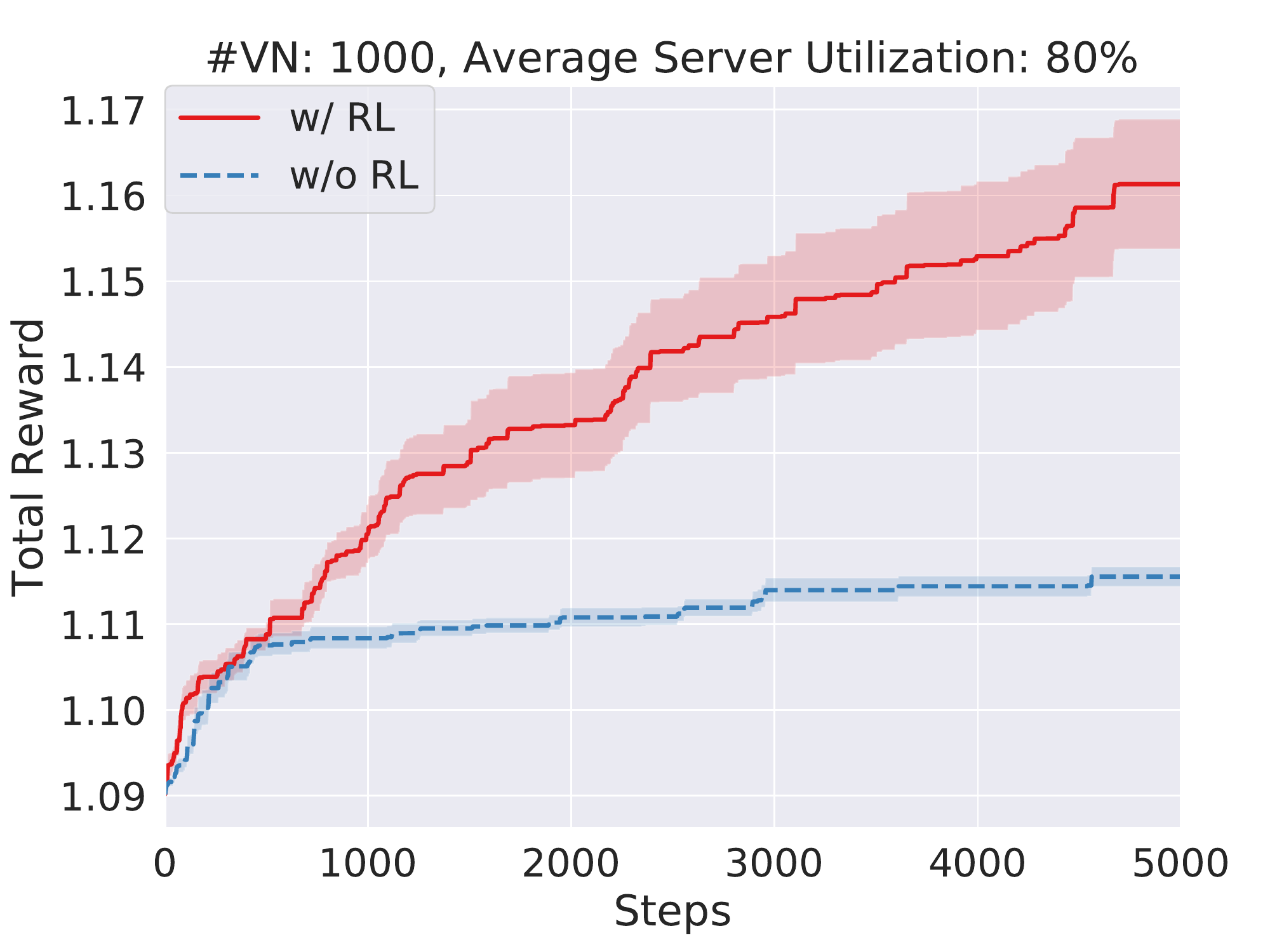}
    \label{fig:1000tr}
}
\subfigure[$N_{\rm VN} = 1500$]{
    \includegraphics[width=0.235\linewidth]{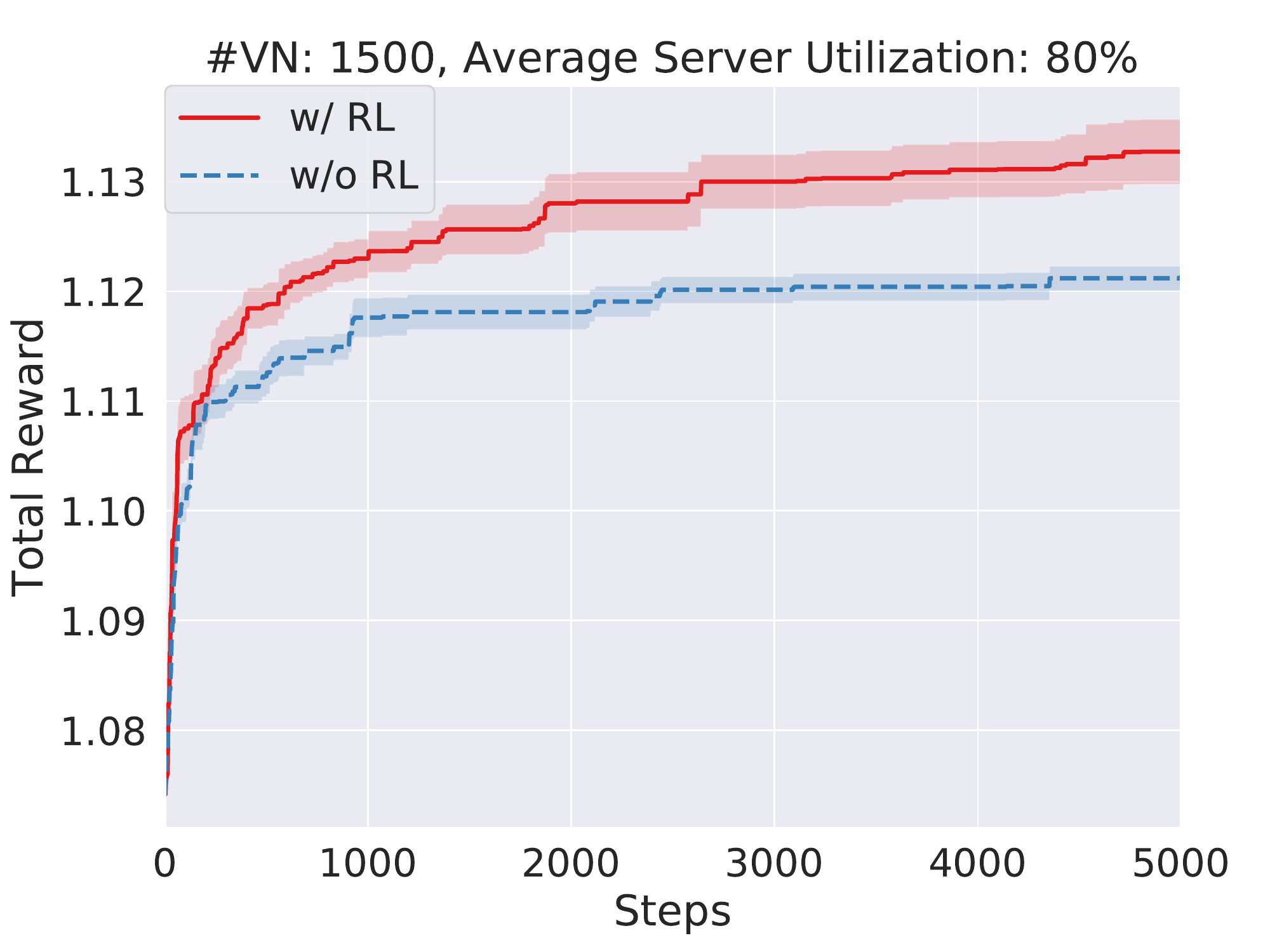}
    \label{fig:1500tr}
}
\subfigure[$N_{\rm VN} = 2000$]{
    \includegraphics[width=0.235\linewidth]{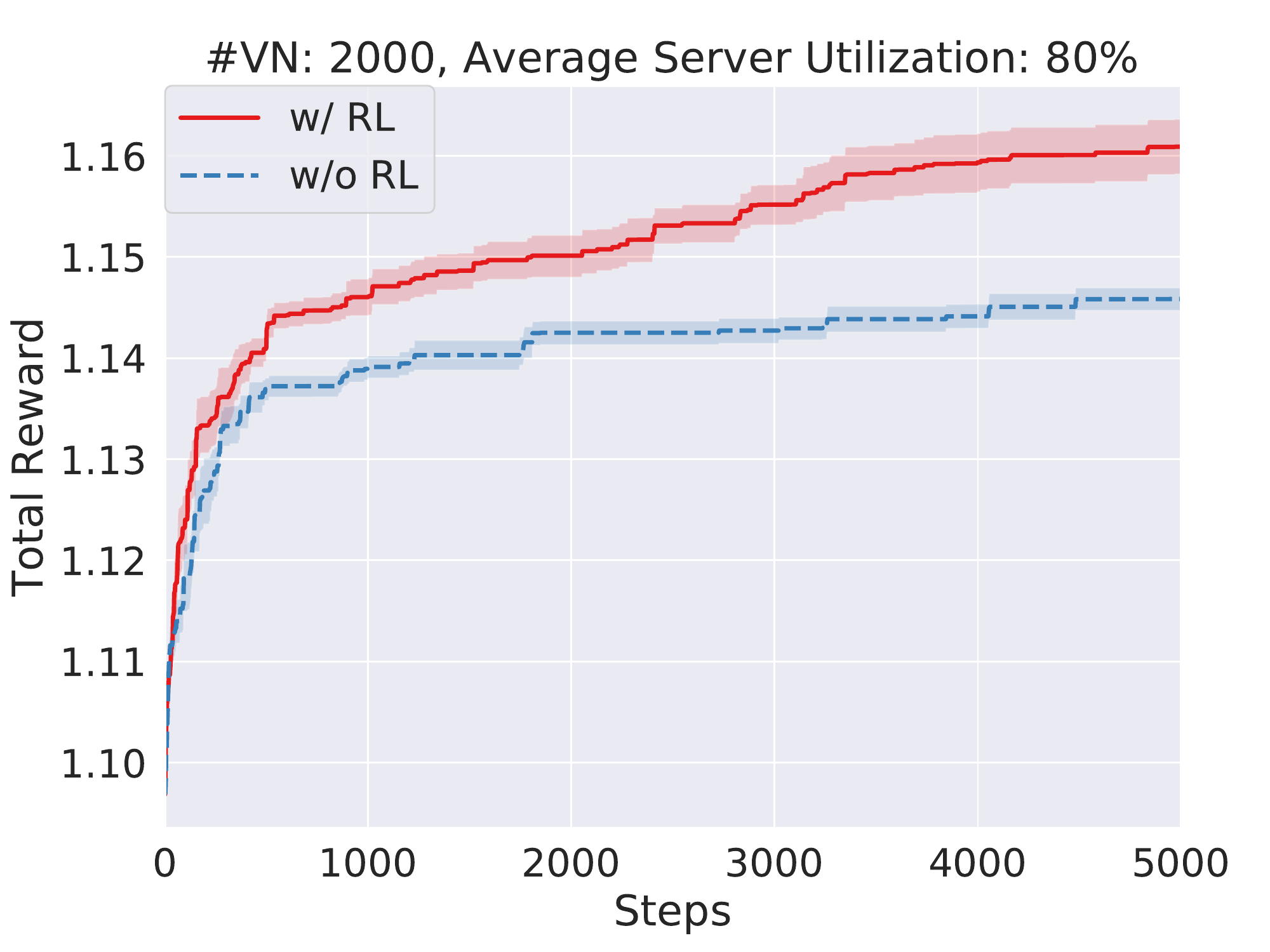}
    \label{fig:2000tr}
}
\caption{Solution-exploration speed for Case \#$1$ and its $N_{\rm VN}$ dependency}
\label{fig:scale_tr}
\end{figure*}
\begin{figure*}[!t]
\centering
\subfigure[$N_{\rm VN} = 20$]{
    \includegraphics[width=0.235\linewidth]{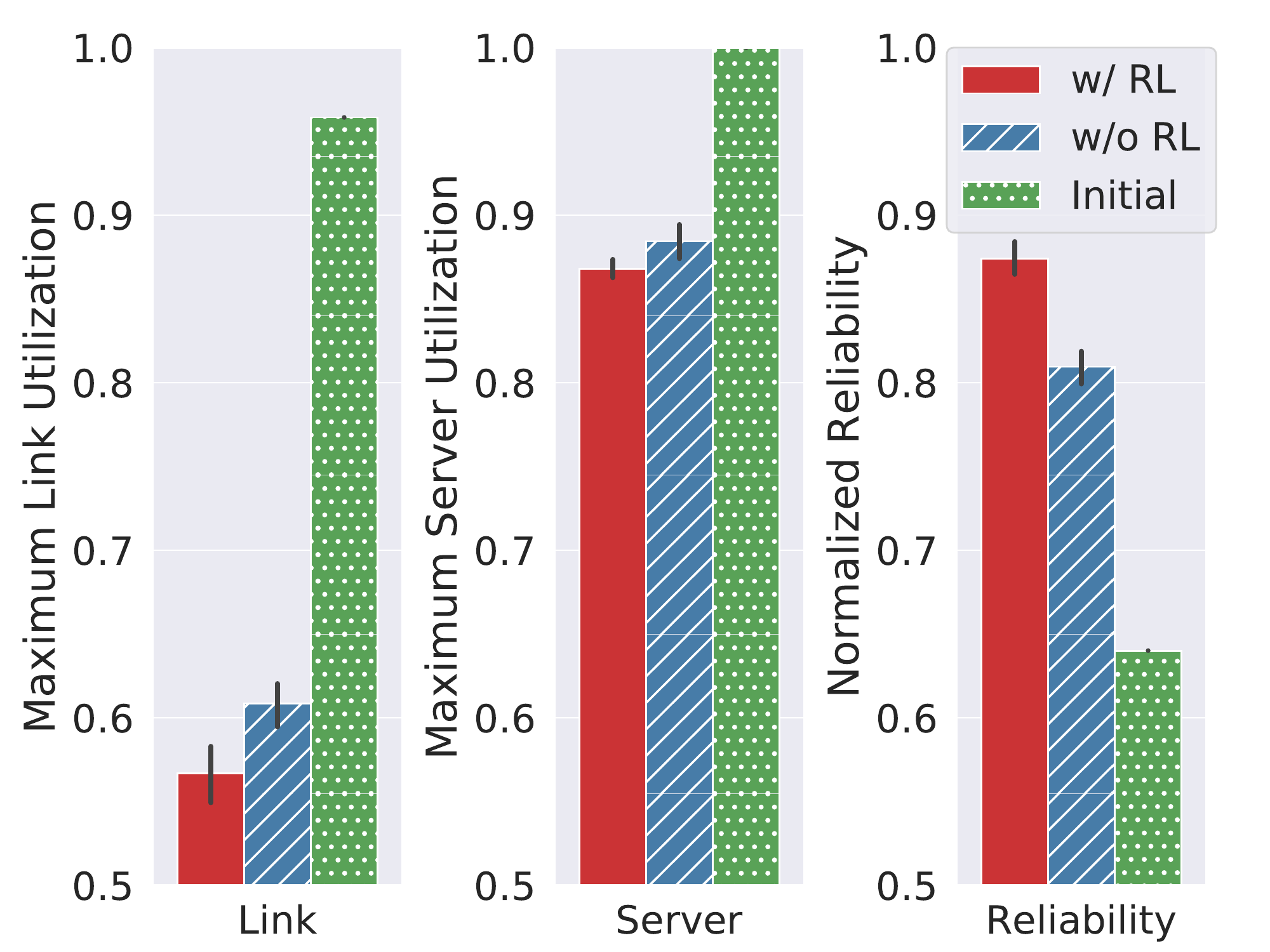}
    \label{fig:20re}
}
\subfigure[$N_{\rm VN} = 50$]{
    \includegraphics[width=0.235\linewidth]{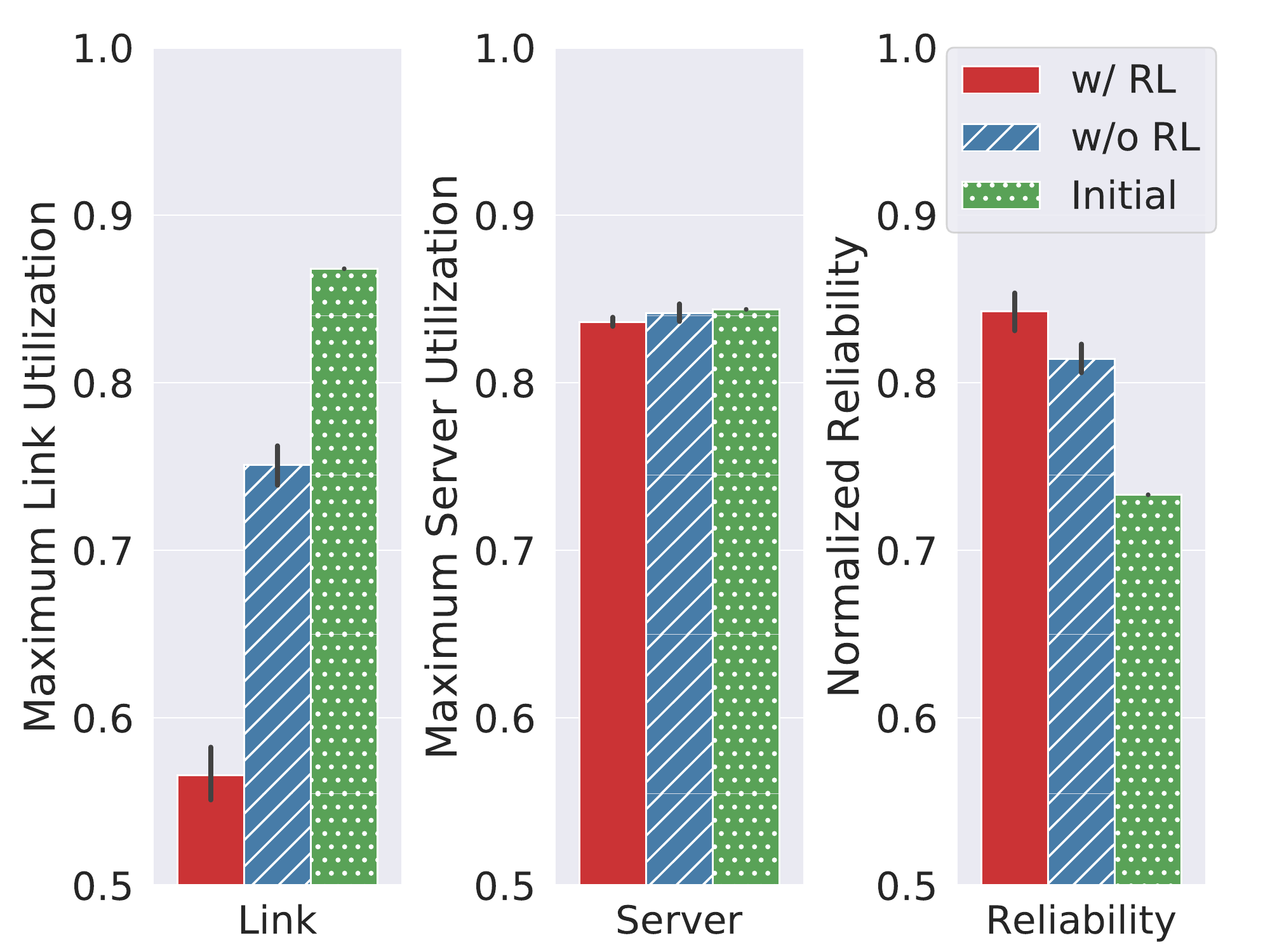}
    \label{fig:50re}
}
\subfigure[$N_{\rm VN} = 200$]{
    \includegraphics[width=0.235\linewidth]{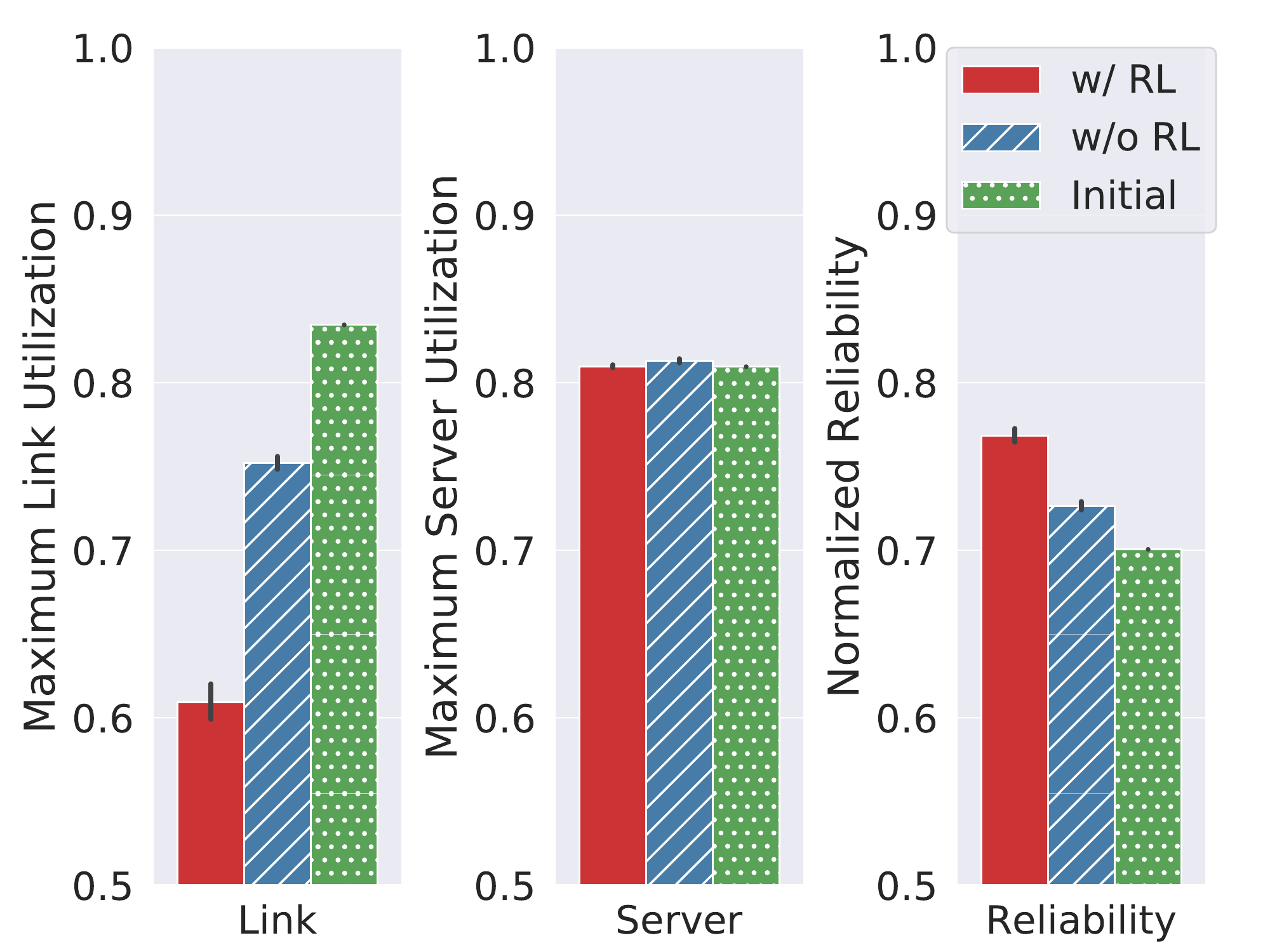}
    \label{fig:200re}
}
\subfigure[$N_{\rm VN} = 400$]{
    \includegraphics[width=0.235\linewidth]{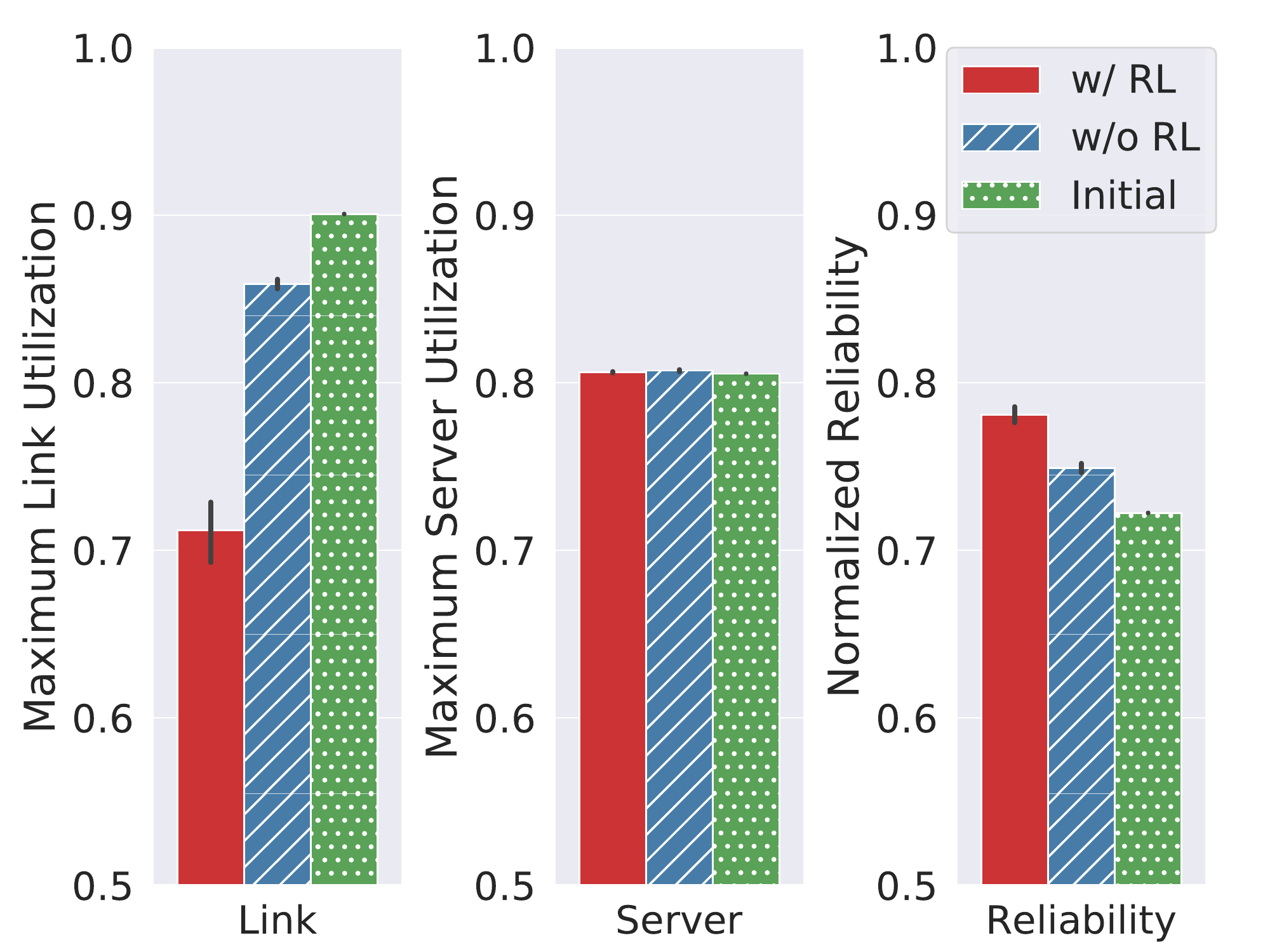}
    \label{fig:400re}
}
\subfigure[$N_{\rm VN} = 800$]{
    \includegraphics[width=0.235\linewidth]{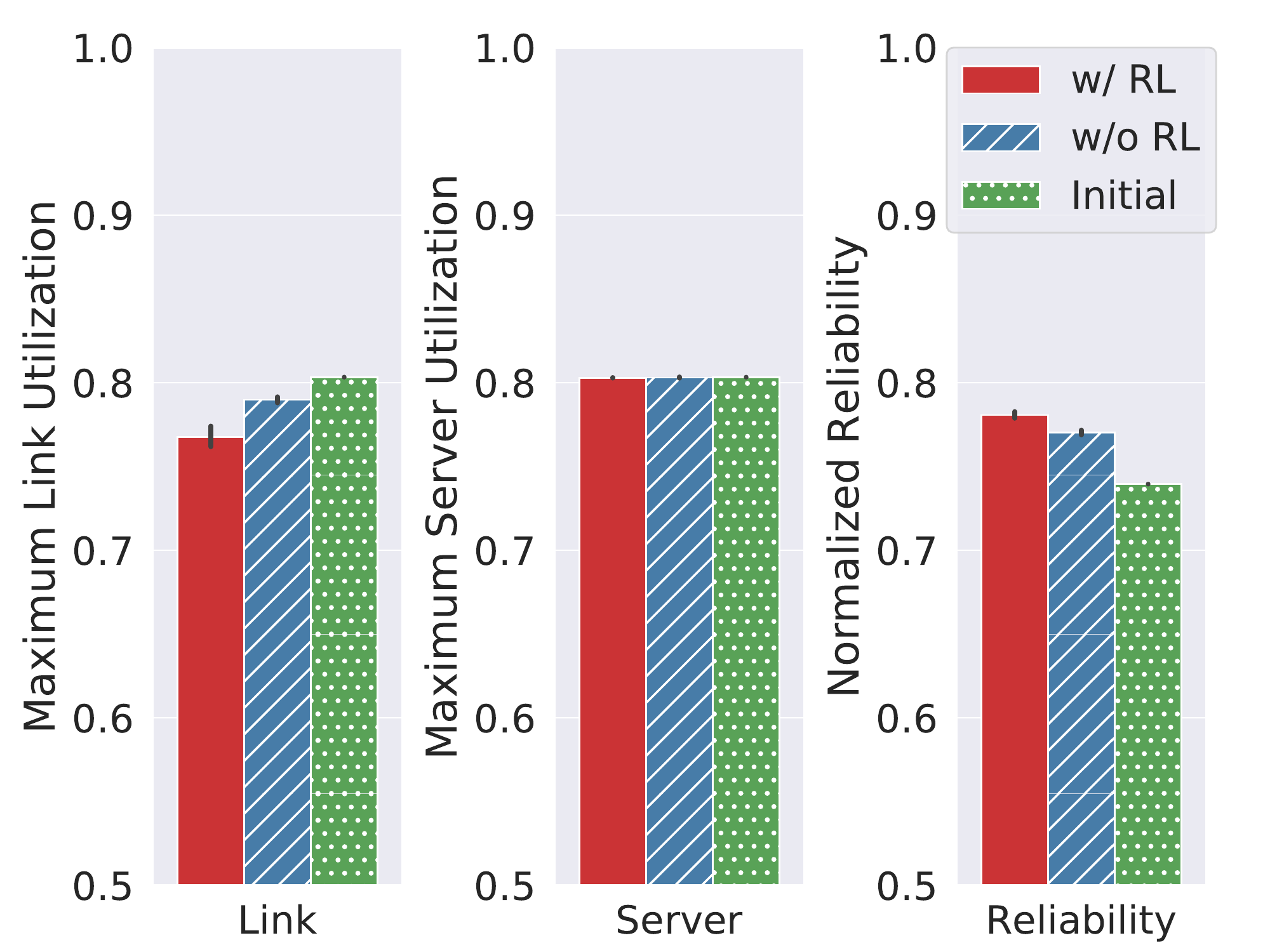}
    \label{fig:800re}
}
\subfigure[$N_{\rm VN} = 1000$]{
    \includegraphics[width=0.235\linewidth]{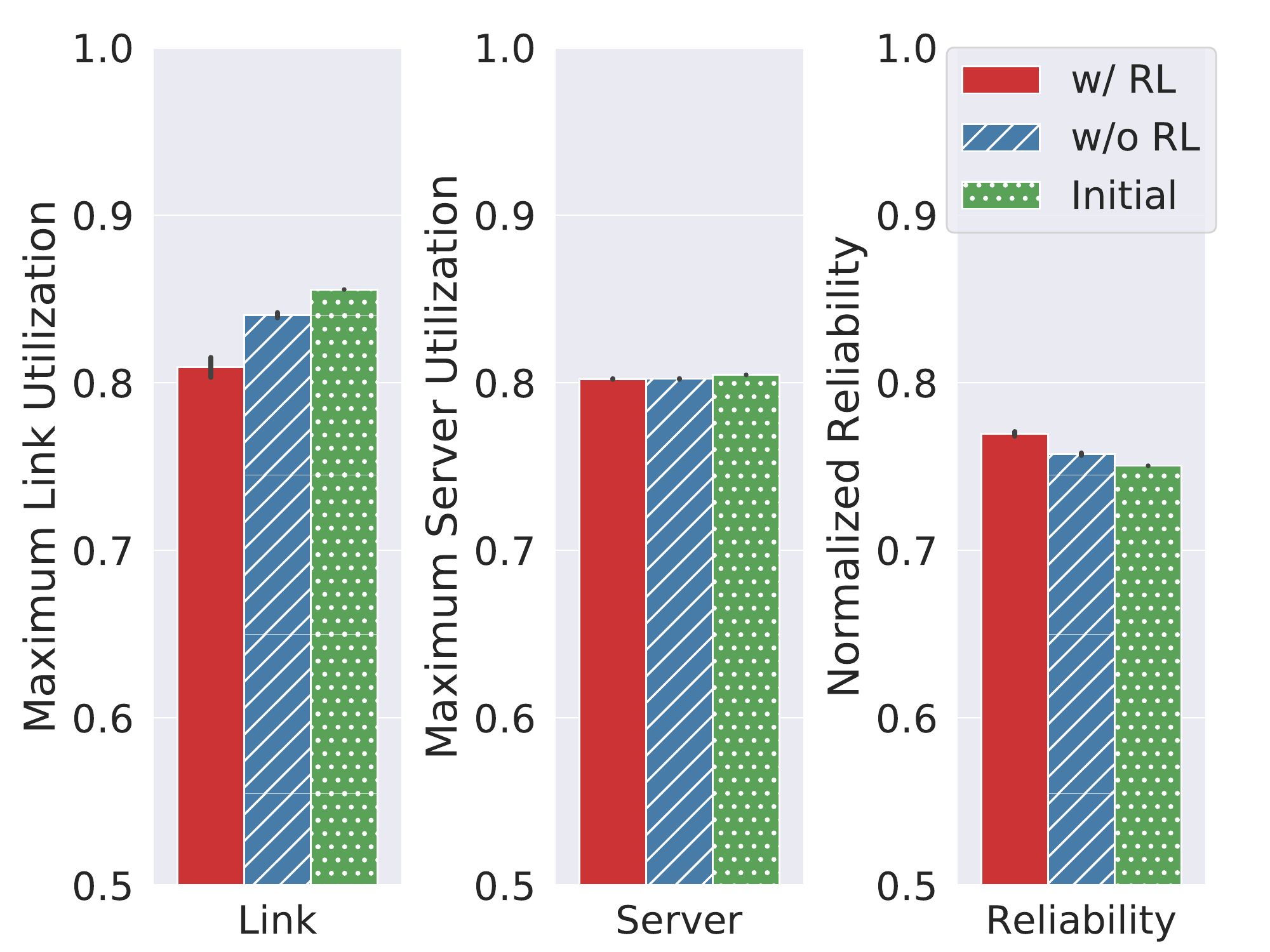}
    \label{fig:1000re}
}
\subfigure[$N_{\rm VN} = 1500$]{
    \includegraphics[width=0.235\linewidth]{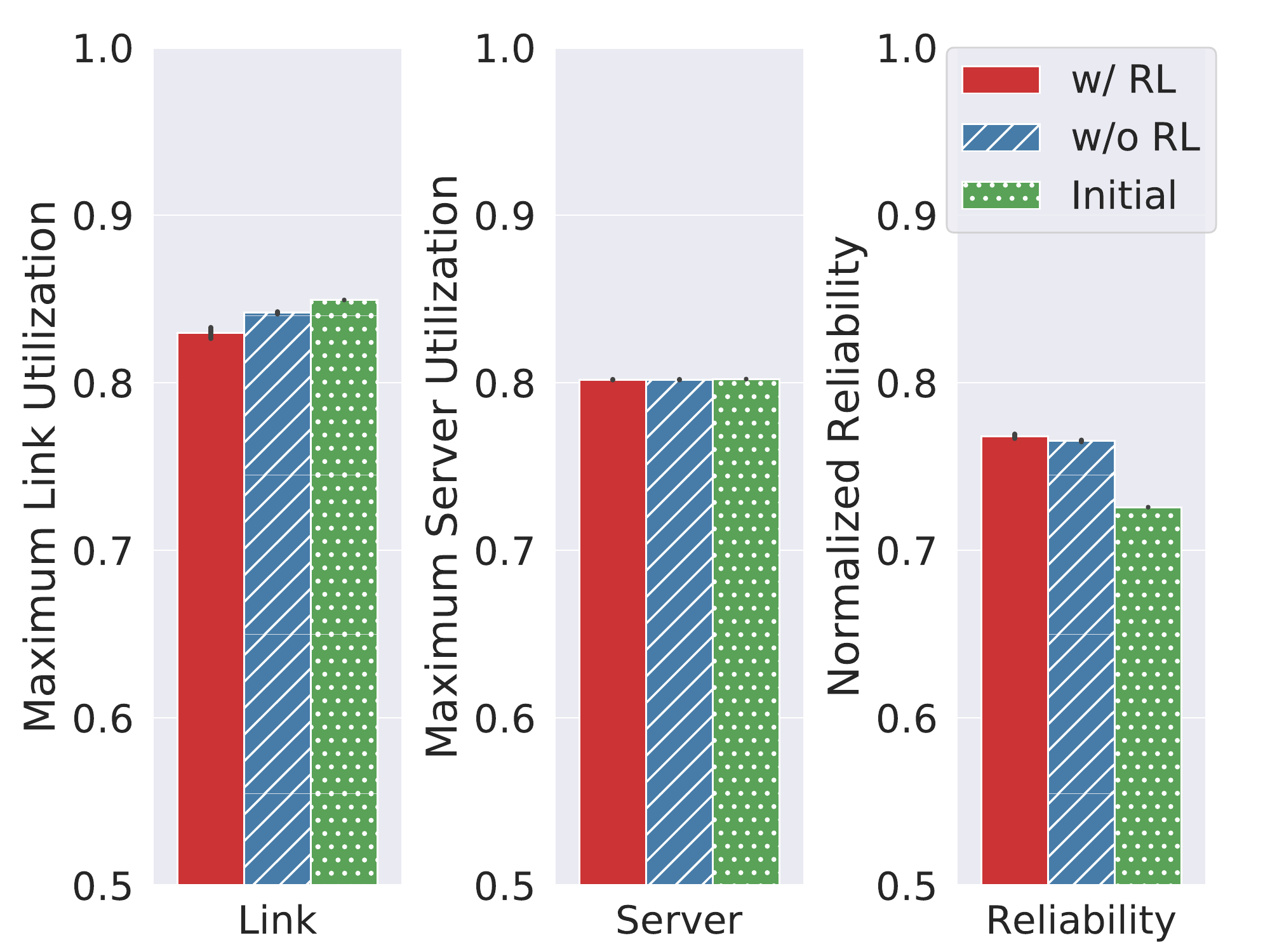}
    \label{fig:1500re}
}
\subfigure[$N_{\rm VN} = 2000$]{
    \includegraphics[width=0.235\linewidth]{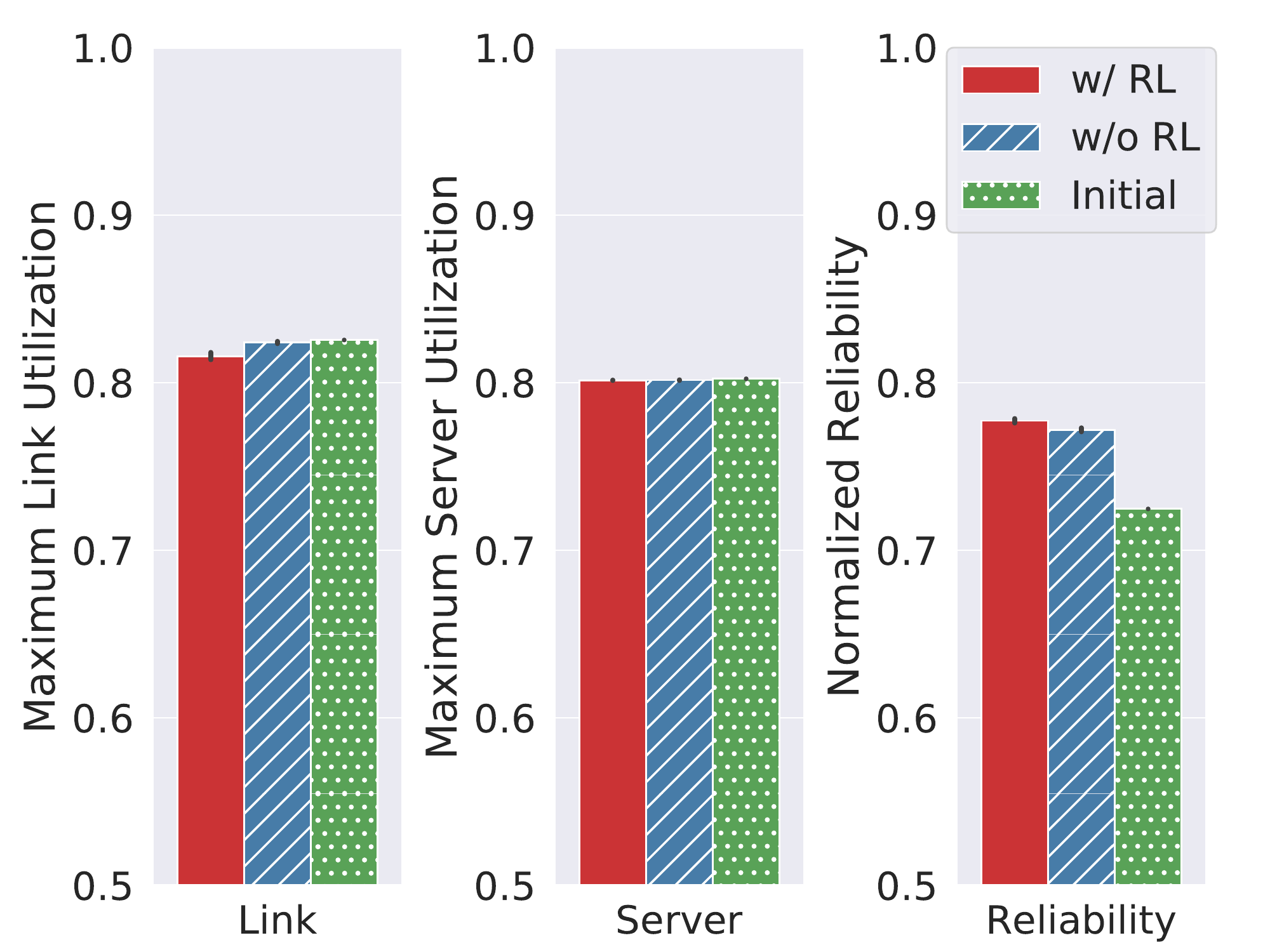}
    \label{fig:2000re}
}
\caption{Components of each objective function value in the best solution for Case \#$1$ and its $N_{\rm VN}$ dependency}
\label{fig:scale_re}
\end{figure*}

\subsubsection{Discussion on solution-exploration speed}
We compared the solution-exploration speeds of the proposed algorithm based on RL (w/ RL) and an algorithm based on changing solutions randomly (w/o RL). Note that, in the case of w/o RL, we set $\epsilon = 1$ and also skipped both agent learning steps, i.e., lines $7$--$8$ in Algorithm~\ref{alg1} and lines $10$--$11$ in Algorithm~\ref{alg2}. In this evaluation, we use Case \#$1$ and set to each weighting parameter $\theta^{\rm link}=\theta^{\rm server}=\theta^{\rm r}=1$.

We first discuss the case when $N^{\rm VN} = 200$ as a baseline. We will discuss other figures~\ref{fig:20tr}--\ref{fig:2000tr} in Sections~\ref{sec:conv} and \ref{sec:scale}. Figure~\ref{fig:200tr} shows the solution-exploration speeds for Case \#$1$, which is the average transition of the best CEV, which is defined by the highest CEV found until the current exploring step. Note that each time CEV is defined by the total reward $r^{g}_t$ shown in (\ref{eq_sogo}). We carried out $10$ calculations with a fixed initial solution. The width of each line indicates the standard deviation ($\pm \sigma$). Though the initial CEV was low due to the control conflict mentioned in Section~\ref{motivation}, the best CEV was improved by repeating the exploration in both cases w/ RL and w/o RL. Results of the comparison between w/ and w/o RL in Fig.~\ref{fig:200tr} indicated that RL could find a better CEV solution within 5000 exploration steps. The reason is that the agent of RL learns the strategy for how to find better allocations efficiently from past exploration steps.

Figure~\ref{fig:200re} shows the components of each objective function value in the best solution for Case \#$1$. Initial in Fig.~\ref{fig:200re} means above values for the initial solution. w/ RL improves the total reward about $0.3$ in Fig.~\ref{fig:200tr}, which is equivalent to improving the sum of link utilization, server utilization, and reliability by $30\%$. Since we assumed $\theta^{\rm link}=\theta^{\rm server}=\theta^{\rm r}=1$ in this evaluation, an increase of 0.01 for the CEV is equivalent to a 1\% improvement in the sum of link utilization, server utilization, and reliability. Of the $30\%$ total improvement, the maximum link utilization improvement is about $22\%$ and the total reliability improvement is about $8\%$. Note that, since the average server utilization is set to $80\%$ in all evaluations, the optimal value of the maximum server utilization is $80\%$. In addition, since an initial solution of VM and IDS placements is calculated by the VM and IDS control engines for minimizing maximum server utilization, this value of the initial solution is near to $80\%$.

\subsubsection{Discussion on difference to optimal solution}\label{sec:conv}
We discuss the difference to the optimal solution in the case of w/ RL and Case $\#1$. General VN allocation problems are known to be NP-hard~\cite{amaldi2016computational}. In addition, there are no previous studies for calculating an optimal solution of Case $\#1$ without approximation. On the other hand, the policy value $Q(s, a)$ of RL has been analytically proved to converge to the optimal policy $Q^{*}(s, a)$ in an infinite number of exploration steps by a policy improvement theorem~\cite{jaakkola1994convergence}. Therefore, when increasing the total exploration steps $T$ to infinity, the solution and its CEV absolutely converge to the optimal solution and optimal value. Since infinite iterations are impossible, we regarded a sub-optimal solution/CEV as the converged solution/CEV when the number of exploration steps sufficiently increased.

Table~\ref{table_opt} shows the convergence speed to the sub-optimal CEV when $N_{\rm NV} = 200$. This evaluation corresponds to the case where the number of exploration steps was increased for the evaluation in Fig.~\ref{fig:200tr}. Since the best CEV sufficiently converges when the total exploration steps $T$ are increased to $1.5 \times 10^6$, CEV = 1.58 is regarded as the sub-optimal CEV and the solution at the time is regarded as the sub-optimal solution. The convergence ratio is defined as the best CEV minus initial CEV divided by sub-optimal CEV minus initial CEV. We defined sufficient converge as the case when the error of the convergence ratio has converged to $1\%$ or less. As shown in Table~\ref{table_opt}, the convergence ratio reaches $56\%$ of the sub-optimal solution in 5000 steps and $82\%$ of the sub-optimal solution in 50,000 steps. This convergence speed seems to suffice as a general NP-hard problem solution.

\begin{table}[!t]
\renewcommand{\arraystretch}{1.3}
\caption{CEV convergence ratio when w/~RL and $N_{\rm VN} = 200$}
\label{table_opt}
\centering
\begin{tabular}{l|c|c}
\hline
{\bf Steps} & Best CEV & Convergence ratio \\ 
\hline
\hline
$0$ (Initial)	& 1.06 & 0.00 \\
$5.0 \times 10^3$	& 1.35 & 0.56 \\
$1.0 \times 10^4$	& 1.38 & 0.62 \\
$5.0 \times 10^4$	& 1.48 & 0.82 \\
$1.0 \times 10^5$	& 1.52 & 0.89 \\ 
$5.0 \times 10^5$	& 1.57 & 0.99 \\
$1.0 \times 10^6$	& 1.58 & 1.00 \\
$1.5 \times 10^6$	& 1.58 & 1.00 \\
\hline
\end{tabular}
\end{table}

\subsubsection{Discussion on scalability}\label{sec:scale}
Figures~\ref{fig:scale_tr} and \ref{fig:scale_re} show the solution-exploration speed and the components of each objective function value in the best solution when we varied $N_{\rm VN}$ from $20$ to $2000$ for Case \#$1$. It reveals that our method can improve the solution by repeating the exploration in all cases of both algorithms (w/ RL and w/o RL). In addition, RL can more efficiently explore better solutions than w/o RL. Note that the performance of the initial solution depends on the randomness of the initial OD traffic and initial client node, so it cannot be compared uniformly in each case.

In the case of $N_{\rm VN}$ is $200$, the improvement of the solution is about $30\%$ for w/ RL and about $10\%$ for w/o RL. On the other hand, when the $N_{\rm VN} = 20$ and $N_{\rm VN} = 1500$ or more, the improvement of the solution is reduced to $10\%$ or less. This shows that the improvement of the solution basically decreases as the $N_{\rm VN}$ increases except for the case of $N_{\rm VN} = 20$. The performance decreased in the case of $N_{\rm VN} = 20$ because a better solution was found easily even w/o RL since the solution exploration space is sufficiently small. The performance decreased in the case of $N_{\rm VN} = 1500$ or more because the learning of RL is not sufficient due to the number of total exploration steps close to the $N_{\rm VN}$. From the above discussion, we conclude that our proposed method was effective in the range of $N_{\rm VN} = 50$ to $1000$.

\begin{figure}[!t]
\centering
\includegraphics[width=0.95\linewidth]{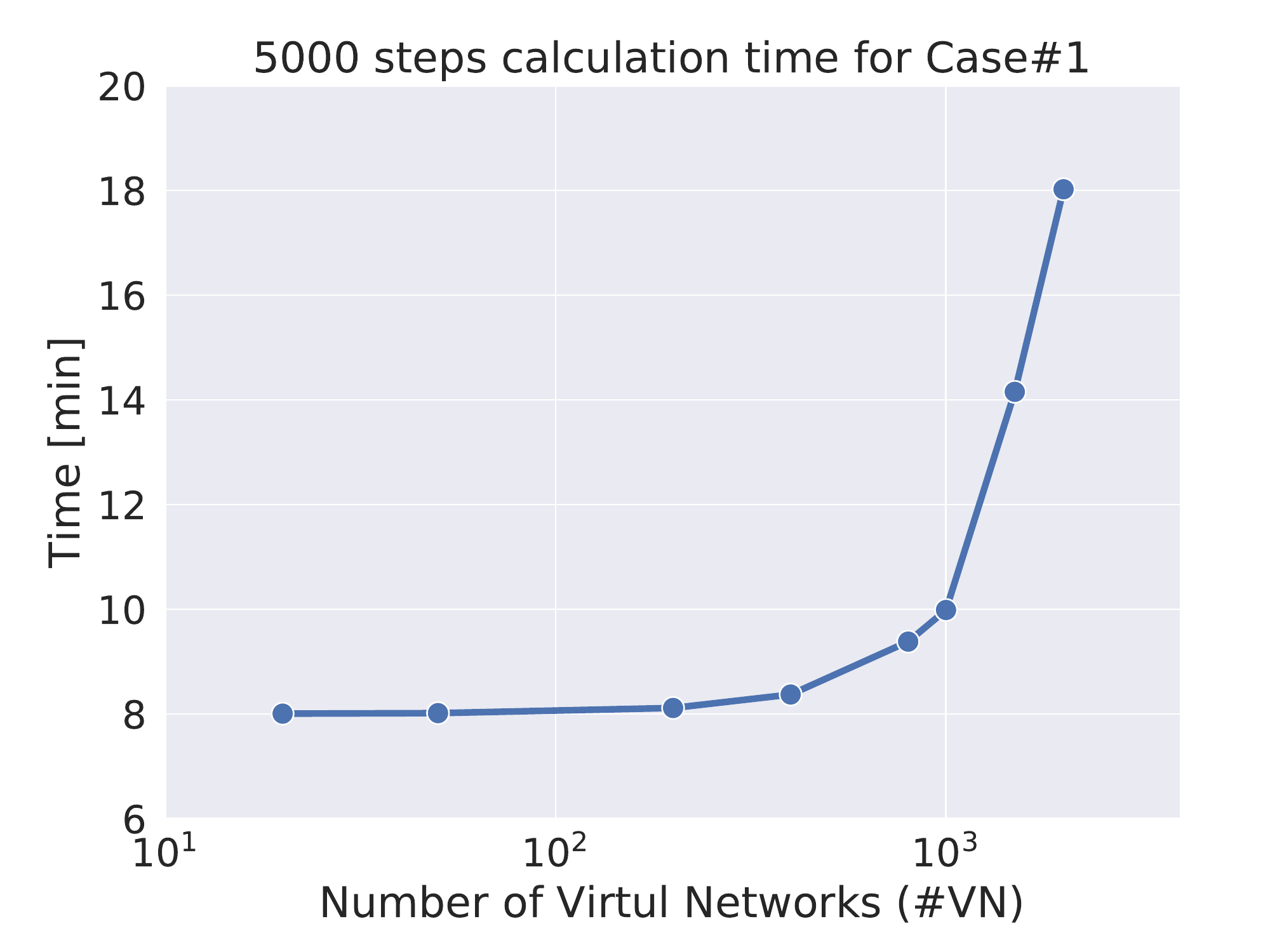}
\caption{Computation time}
\label{fig:time}
\end{figure}

Figure~\ref{fig:time} shows the computation time of the proposed algorithm (w/ RL) for $N_{\rm VN}$~$20$ and $2000$. The calculations were performed on a Intel core i$7$ $4790$k CPU of a single core. The computation time increases depending on the number of steps proportionally. Although $N_{\rm VN}$ increased $100$ times, the computation time increased only several times. This means that, from the viewpoint of computation time, the proposed method is scalable with respect to $N_{\rm VN}$, with up to $1000$ VNs. Note that w/o RL has almost the same computation time as w/ RL. The difference between w/ RL and w/o RL is the overhead time of RL, which is less than $1\%$ of the total computation time.

Figure~\ref{fig:time} also shows that the calculation time of the proposed method until $5000$ steps is less than 10 minutes in the range up to $1000$ VNs. We thus considered that the calculation time allows enough practice. In NFV environments, each VN demand is statistically multiplexed by multiple users sharing the VN. For that reason, our proposed control system mainly targets static VN demand allocation, which considers VN demands to be fixed within a particular period (e.g., more than 1 hour). Moreover, our method can adjust the calculation time to modify the number of total exploration steps in accordance with the required calculation time.

The computation time of our method is determined by the calculation time of each control engine. In this evaluation, the route control engine is formulated as a linear programming (LP) problem, and its calculation time is less than $1$ second. However, if the route control engine is formulated as an integer linear programming (ILP) problem, e.g., non-split route case and path-base route control case, the computation time of our method will increase dramatically. When each exploration step contains ILP problems, the following solution seems to be effective: set the upper limit for calculation time of each step, approximate by limiting of route candidates, and use the heuristic method for route calculation.

\begin{table*}[!t]
\renewcommand{\arraystretch}{1.3}
\caption{Parameters depend on each case when $N_{\rm VN}$ is $200$}
\label{table_ext}
\centering
\begin{tabular}{ll||c|c|c|c|c|c|c|c|c|c|c|c}
\hline
{\bf Definitions} & & 1 & 2 & 3 & 4 & 5 & 6 & 7 & 8 & 9 & 10 & 11 & 12 \\ 
\hline
\hline
Number of IDSs & $N_{\rm ids}$          & 200 & 10 & 200 & 10 & 200 & 10 & 200 & 10 & 0 & 0 & 0 & 0 \\
Number of clients & $N_{\rm cli}$       & 200 & 200 & 0 & 0 & 200 & 200 & 0 & 0 & 200 & 0 & 200 & 0 \\
$i^{\rm th}$ VM size & $w_{i}^{\rm vm}$       & 1--3 & 1--3 & 1--3 & 1--3 & 1--3 & 1--3 & 1--3 & 1--3 & 3--8 & 3--8 & 3--8 & 3--8 \\
$i^{\rm th}$ IDS size & $w_{i}^{\rm ids}$     & 2 & 40 & 2 & 40 & 2 & 40 & 2 & 40 & 0 & 0 & 0 & 0 \\
$i^{\rm th}$ IDS capacity & $c_i^{\rm ids}$   & 1 & 20 & 1 & 20 & 1 & 20 & 1 & 20 & 0 & 0 & 0 & 0 \\
\hline
\end{tabular}
\end{table*}

\begin{figure*}[!t]
\centering
\subfigure[Case~\#$1$ (w/ IDS, w/ Reliability, w/ Fixed node, IDS isolation)]{
    \includegraphics[width=0.32\linewidth]{fig/case1_w111_vn200_result.pdf}
    \label{fig:case1}
}
\subfigure[Case~\#$3$ (w/ IDS, w/ Reliability, IDS isolation)]{
    \includegraphics[width=0.32\linewidth]{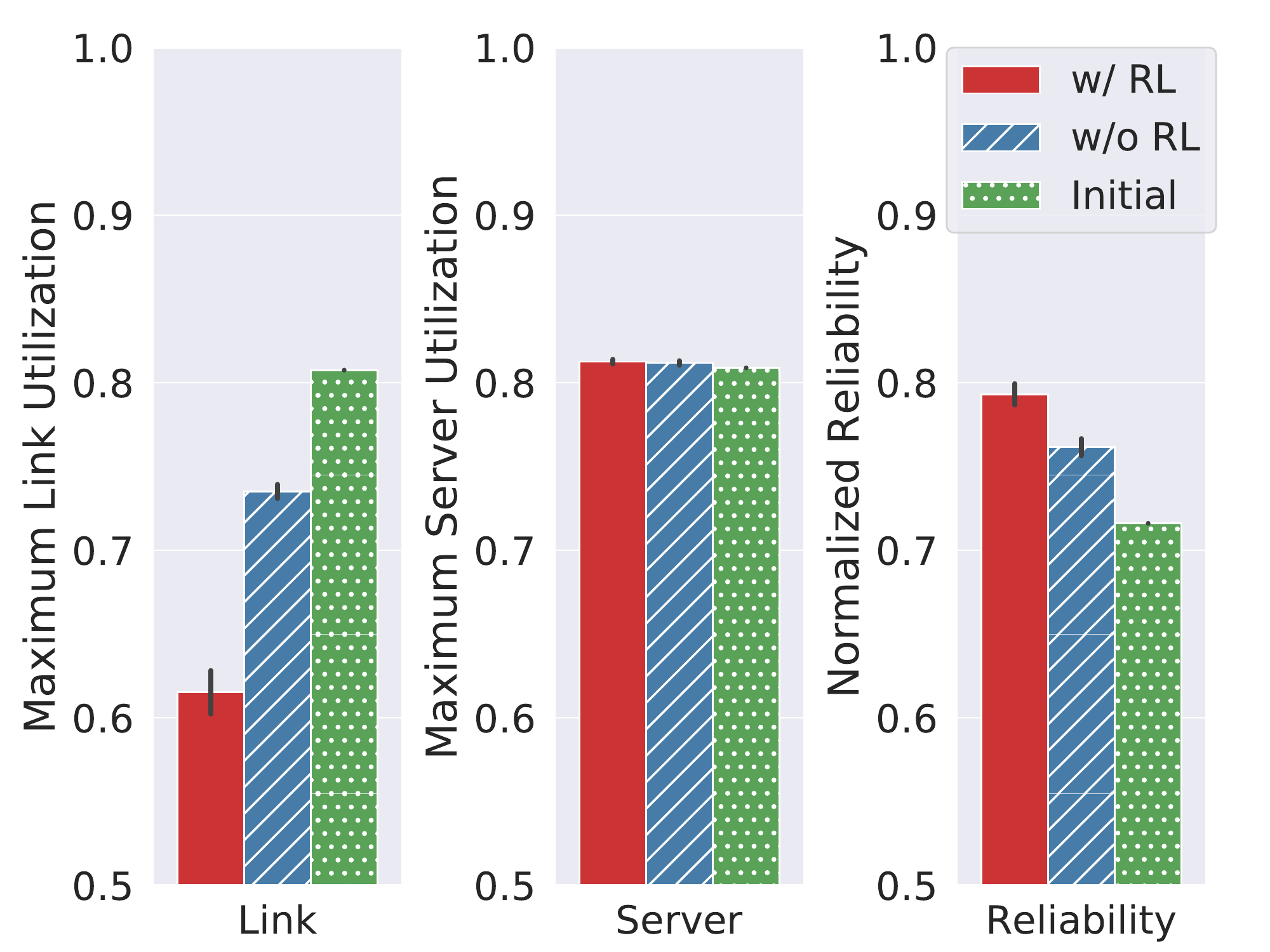}
    \label{fig:case3}
}
\subfigure[Case~\#$4$ (w/ IDS, w/ Reliability, IDS sharing)]{
    \includegraphics[width=0.32\linewidth]{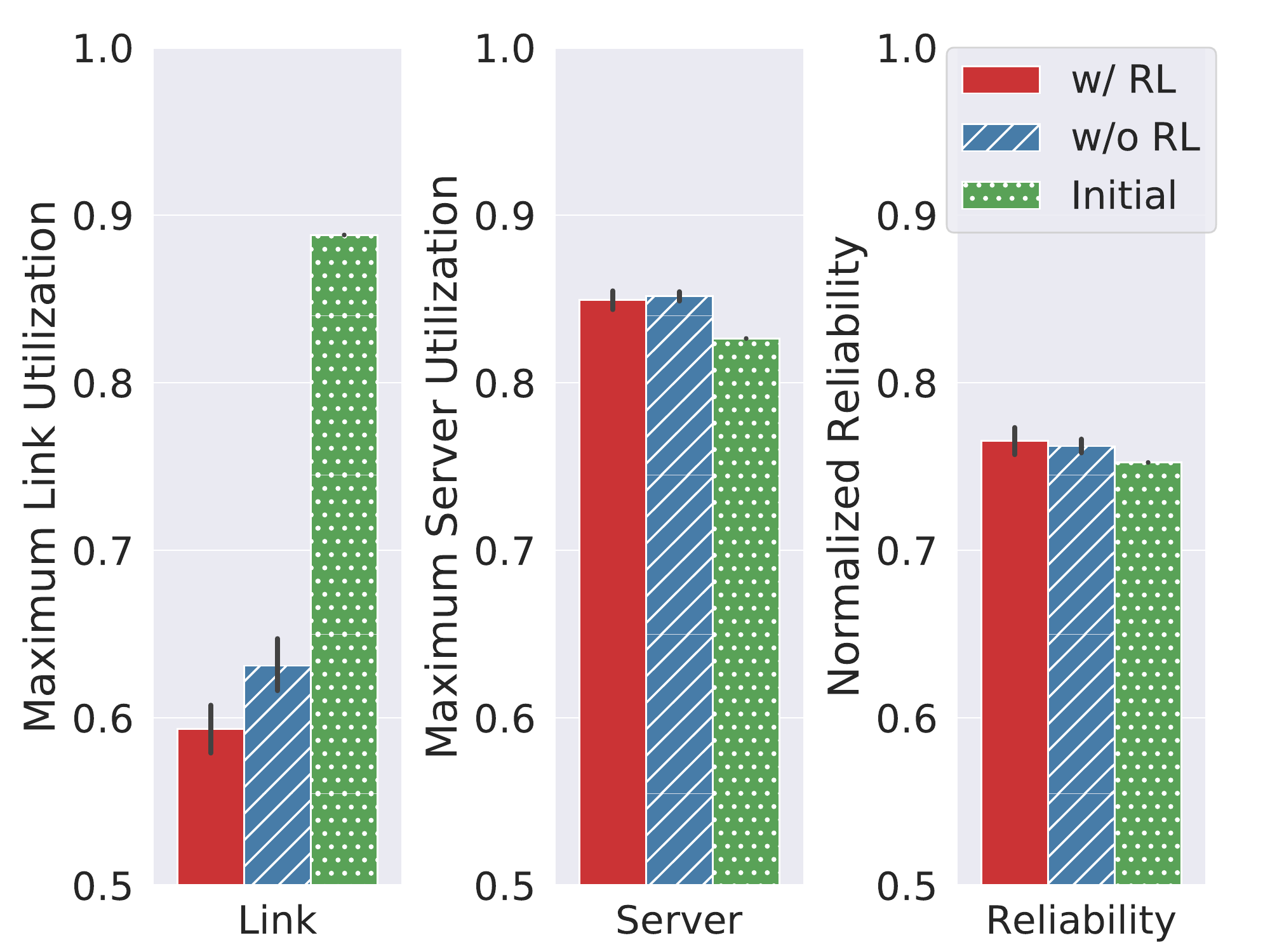}
    \label{fig:case4_w111}
}
\subfigure[Case~\#$5$ (w/ IDS, w/ Fixed node, IDS isolation)]{
    \includegraphics[width=0.32\linewidth]{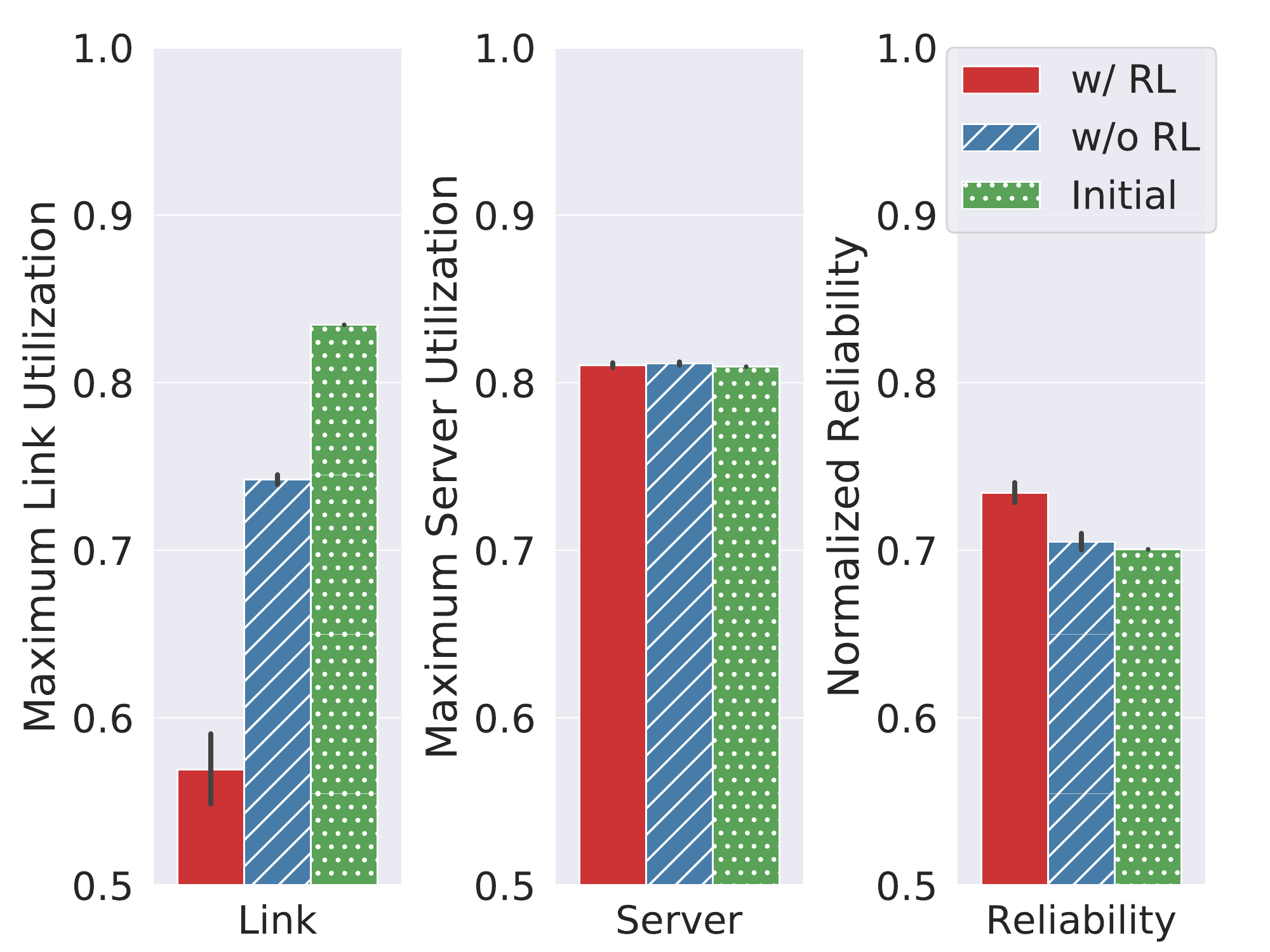}
    \label{fig:case5}
}
\subfigure[Case~\#$8$ (w/ IDS, IDS sharing)]{
    \includegraphics[width=0.32\linewidth]{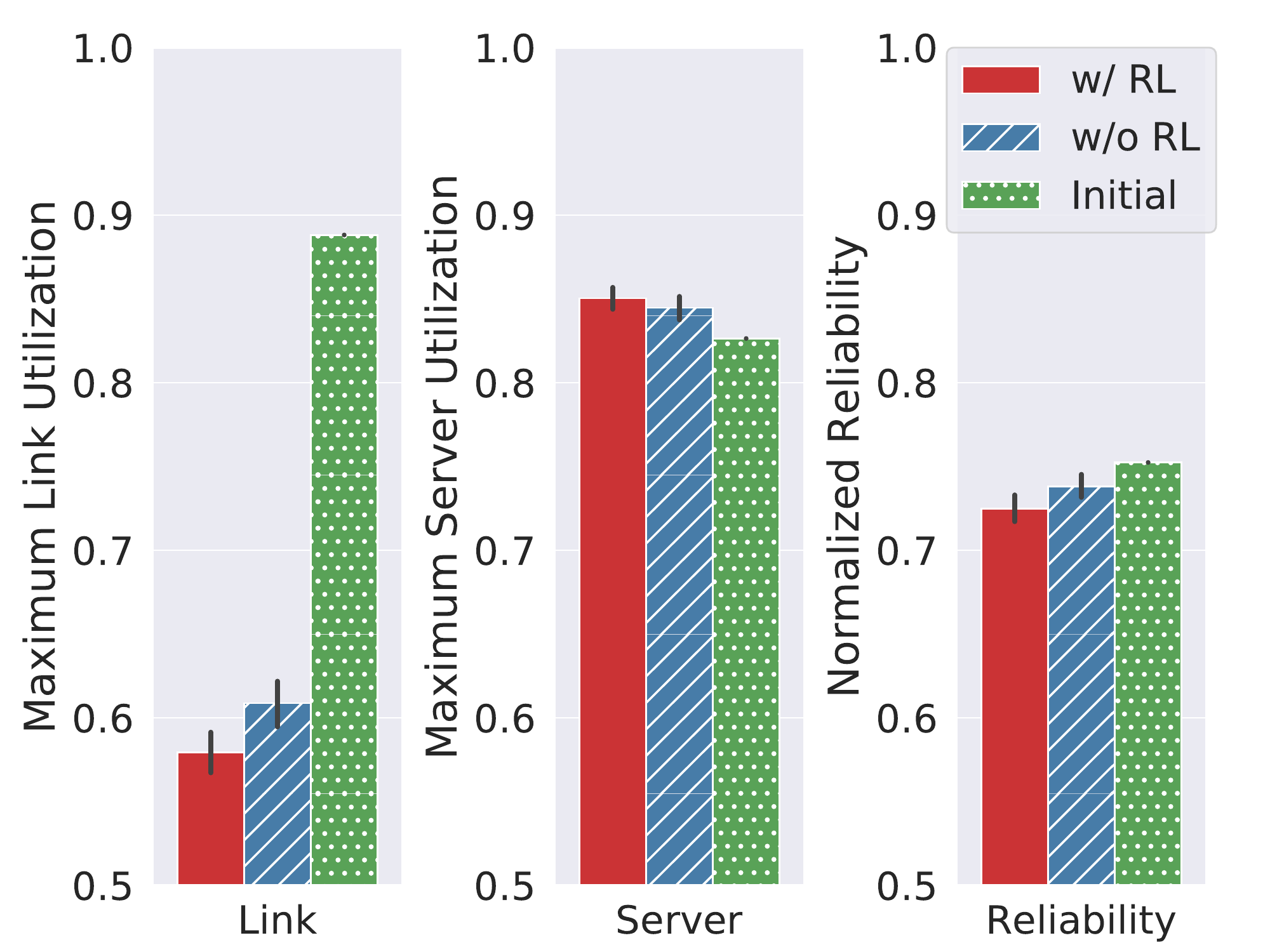}
    \label{fig:case8}
}
\subfigure[Case~\#$12$ (default)]{
    \includegraphics[width=0.32\linewidth]{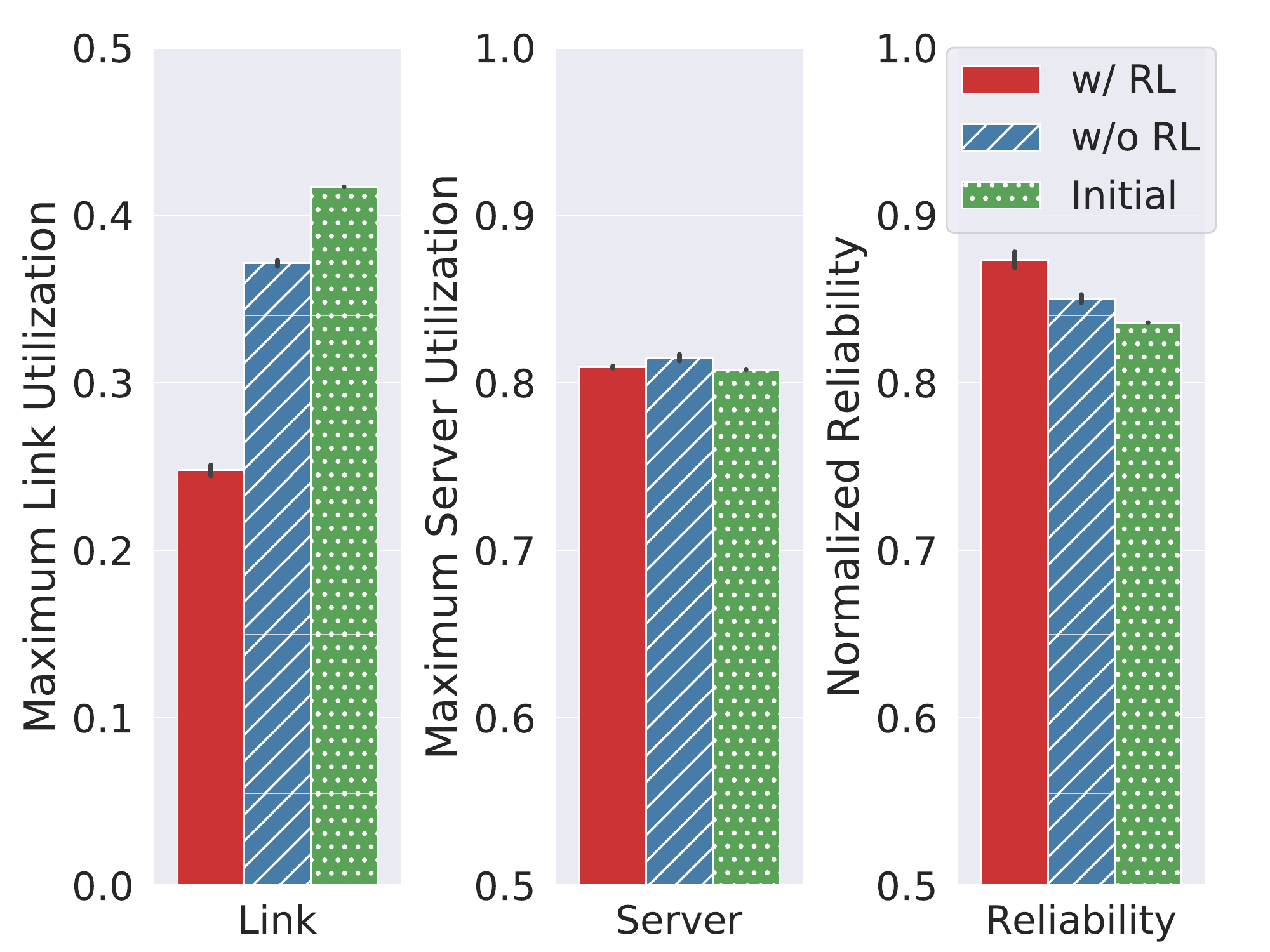}
    \label{fig:case12}
}
\caption{Components of each objective function value in the best solution for each case.}
\label{fig:extend}
\end{figure*}

\begin{figure*}[!t]
\centering
\subfigure[$(\theta^{\rm link}, \theta^{\rm server}, \theta^{\rm r}) = (10, 0, 0)$]{
    \includegraphics[width=0.32\linewidth]{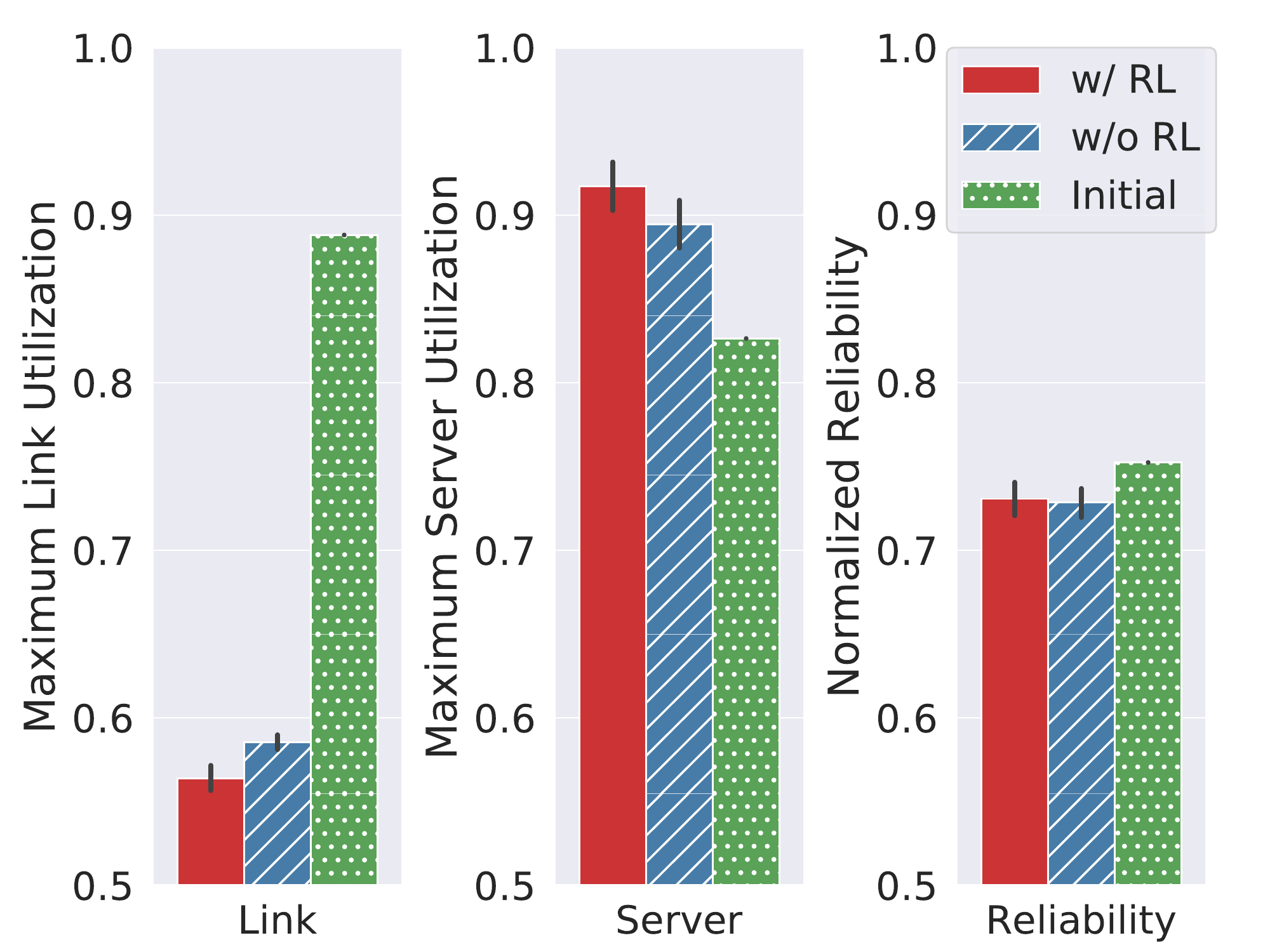}
    \label{fig:case4_w100}
}
\subfigure[$(\theta^{\rm link}, \theta^{\rm server}, \theta^{\rm r}) = (0, 10, 0)$]{
    \includegraphics[width=0.32\linewidth]{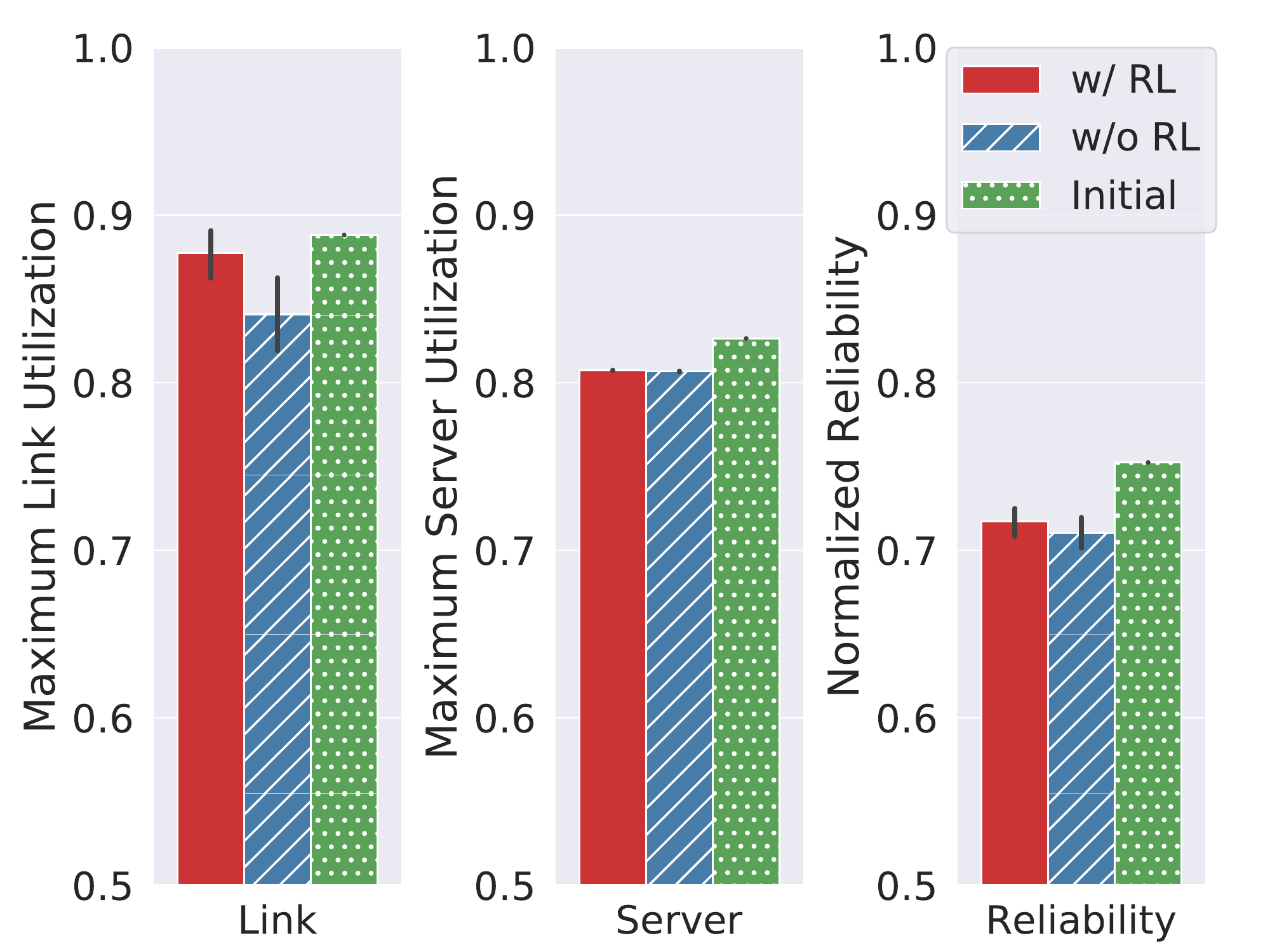}
    \label{fig:case4_w010}
}
\subfigure[$(\theta^{\rm link}, \theta^{\rm server}, \theta^{\rm r}) = (0, 0, 10)$]{
    \includegraphics[width=0.32\linewidth]{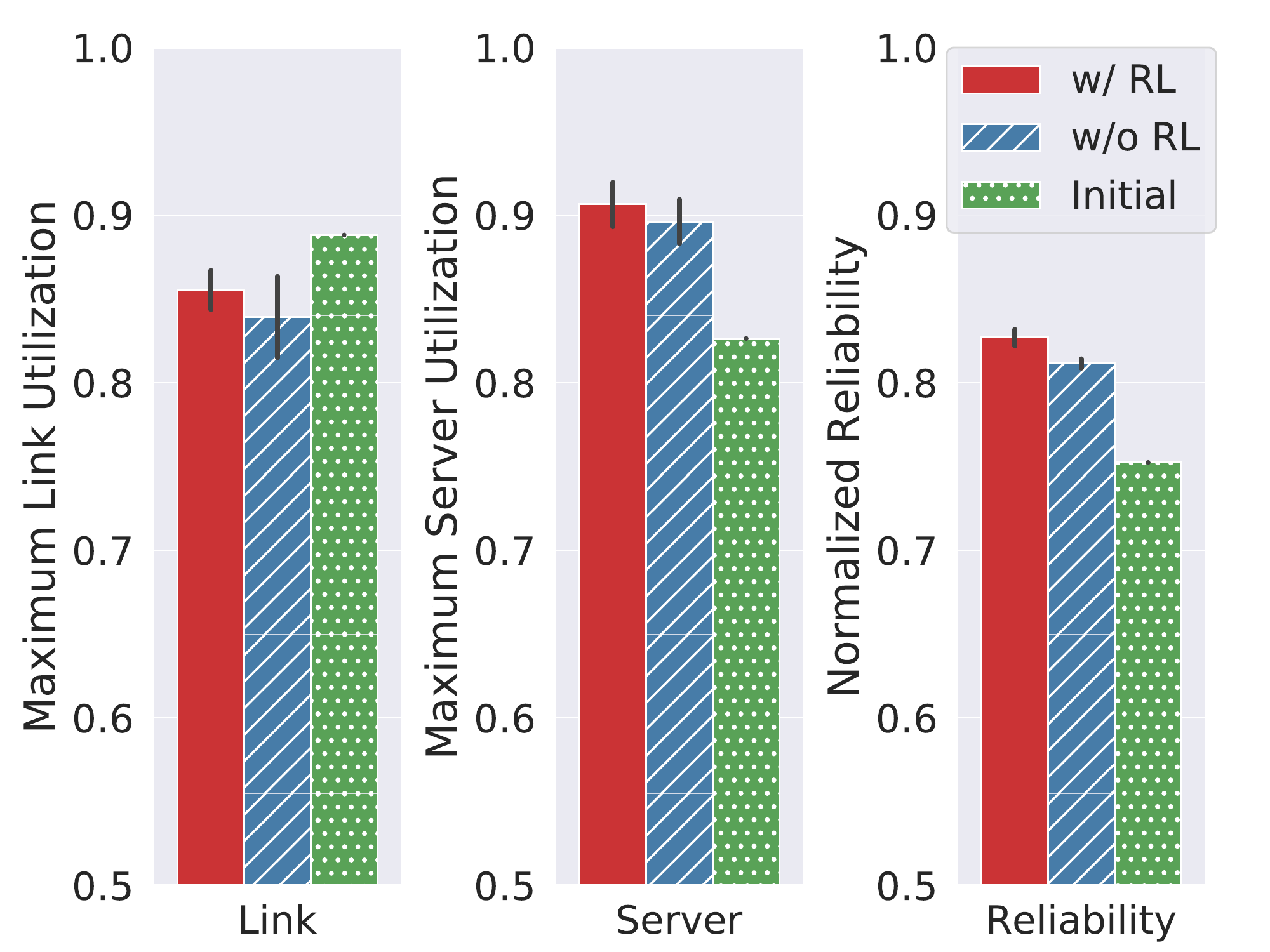}
    \label{fig:case4_w001}
}
\caption{Weighting parameter dependency of components of each objective function value in the best solution for Case \#$4$}
\label{fig:case4_w}
\end{figure*}

\subsubsection{Discussion on extendability}
Since the extendability of our method is difficult to evaluate quantitatively, we evaluated the applicability of the proposed method under the various use cases. Although the applicability is not equal to the extendability, we assume that it indirectly provides evidence that the proposed method has extendability.

Figure~\ref{fig:extend} shows the components of each objective function value in the best solution for each case. In this evaluation, we selected $6$ use cases to discuss the effects of changing each option (Cases \#$1$, \#$3$, \#$4$, \#$5$, \#$8$, \#$12$) and set to each weighting parameter $\theta^{\rm link}=\theta^{\rm server}=\theta^{\rm r}=1$. Results reveal that our method can improve the solution by repeating the exploration in all cases of both algorithms (w/ RL and w/o RL). Therefore, we can indirectly show that our proposed method has highly extendability.

Next, we consider the effects of adding or changing each option in details. First, we describe the result of the simplest Case \#$12$. Figure~\ref{fig:case12} shows that our method improves the maximum link utilization. Although reliability is not considered as an objective function, the total reliability is also improved. It seems that since the solution with a shorter route was preferentially selected to reduce the maximum link utilization, the total reliability was improved coordinately.

We first consider the influence of adding the \textbf{(1) with IDS} option under the IDS sharing option by comparing Figs.~\ref{fig:case8} and \ref{fig:case12}. In Fig.~\ref{fig:case8}, the maximum link utilization of the initial solution is drastically increased by adding IDS. This is because the path length increases due to the addition of the middle server node, and the total traffic volume in a physical network increases proportionally.
We also consider the influence of adding the IDS sharing option.
In Fig.~\ref{fig:case8}, the maximum server utilization is increased and the total reliability is decreased. The reason for increasing maximum server utilization is that large IDSs exist in IDS sharing condition.
This can be seen from the fact that the maximum server utilization does not increase in Cases \#$1$, \#$3$, or \#$5$. The reason the total reliability did not improve as the link utilization improved is that the IDS sharing makes it difficult to find a solution that avoids the disaster area. This difficulty to avoid the disaster area in IDS sharing condition is why the IDS that minimizes the length of the OD route without considering the reliability is preferentially selected.

Second, we consider the influence of adding the \textbf{(2) with Reliability} option by comparing Figs.~\ref{fig:case4_w111} and \ref{fig:case8}. The reliability is clearly improved by maintaining the link utilization efficiency and server utilization efficiency. The comparison between Figs.~\ref{fig:case1} and \ref{fig:case5} shows a similar result.

Third, we consider the influence of changing the \textbf{(3A) with Fixed node} option by comparing Figs.~\ref{fig:case1} and \ref{fig:case3}. Figure~\ref{fig:case1} shows that the total reliability is decreased slightly by introducing fixed clients. This is because the placement of the origin node (i.e., client node) is fixed, which makes it difficult to avoid the disaster area.

Finally, we a consider the influence of changing the \textbf{(3B) IDS isolation or sharing} option by comparing Figs.~\ref{fig:case3} and \ref{fig:case4_w111}. Figure~\ref{fig:case3} shows that the maximum server utilization is decreased and the total reliability is increased by changing the IDS isolation model. The decrease in server utilization efficiency made it easier to improve the server utilization efficiency by removing the large IDSs. The increase in total reliability made it easier to find the solutions that avoid the disaster area by IDSs isolation for each VN. The maximum link utilization in w/o RL also is increased due to the increase in the solution space by increasing the $N_{\rm ids}$.

\subsubsection{Discussion on weight parameters}
Figures~\ref{fig:case4_w111} and \ref{fig:case4_w} show the effectiveness of weight parameters for Case \#$4$. In this evaluation, we set four different conditions: $(\theta^{\rm link}, \theta^{\rm server}, \theta^{\rm r}) = (1, 1, 1)$, $(10, 0, 0)$, $(0, 10, 0)$, $(0, 0, 10)$. Note that our method can satisfy all constraints even if each weighting parameter is $0$ as shown in Fig.~\ref{fig:case4_w}. 
The results in Fig.~\ref{fig:case4_w100} have the best link utilization efficiency, the results in Fig.~\ref{fig:case4_w010} have the best server utilization efficiency, and the results in Fig.~\ref{fig:case4_w001} have the best total reliability. In particular, the maximum server utilization of $0.8$ is a global optimum solution. From the above results, our method can suggest a wide variety of options of solutions by adjusting the weighting parameters~$\theta^{\rm link}$, $\theta^{\rm server}$, $\theta^{\rm r}$.

\subsubsection{Discussion on difference to previous method}
We discuss the differences between the proposed coordinated method and previous combined approaches in terms of how easy/difficult they are to build and their problems are to solve. As described in Section~\ref{sec:intro}, previous combined approaches need specified algorithm to be built that simultaneously solves the combined optimization problem. We formulated and implemented the combined optimization problem solving the VN allocation problem for Case \#12, which is the simplest use case among Cases \#1--12. We also evaluate the differences in the performance of the solution between the proposed method based on the RL and the previous method based on the combined optimization problem.

We first describe the formulation of the combined optimization problem for Case \#12. The conditions and assumptions for Case \#12 have already been described in Section~\ref{sec:mod_option}. Since the control metrics are routes and VM placements, in this case, we need to formulate the route control algorithm and the VM control algorithm and to newly formulate the relational equations between the variables of both algorithms. We use the route control algorithm shown in (2)--(7) and the VM control algorithm shown in (8)--(12).

Since traffic demands between nodes ${\bm T}^{\rm node} := \{ t_{pq} \}$ in (5) are determined by the traffic demands between VMs ${\bm T}^{\rm vm} := \{ t_{ij}^{\rm vm} \}$ and VM placements ${\bm \Xi}^{\rm vm} := \{ {\xi}^{\rm vm}_{ip} \}$, the relational equations between both algorithms can be formulated as follows.
\begin{eqnarray}
{\bm T}^{\rm node} &=& {}^t{\bm \Xi}^{\rm vm} {\bm T}^{\rm vm} {\bm \Xi}^{\rm vm} \nonumber \\
t_{pq} &=& \sum_{i \in {\bm V}} \sum_{j \in {\bm V}} {\xi}_{ip}^{\rm vm} t_{ij}^{\rm vm} {\xi}_{jq}^{\rm vm},
\end{eqnarray}
where $t_{pq}$ shows traffic demands from node $p$ to node $q$, $t_{ij}^{\rm vm}$ shows traffic demands from $i^{\rm th}$ VM to $j^{\rm th}$ VM, and ${\xi}^{\rm vm}_{ip}$ shows VM allocation, which returns $1$ if $i^{\rm th}$ VM is assigned to the $p^{\rm th}$ node; otherwise, $0$. Therefore, the combined optimization problem is formulated with the objective of minimizing $U_{\rm max}^{\rm link} + U_{\rm max}^{\rm server}$ and the constraints (2)--(12) and (18).

Next, we describe the implementation to solve the combined optimization problem. This problem is categorized into quadratically constrained mixed-integer non-linear programming (QC-MINLP). In this paper, we use Pyomo~\cite{hart2011pyomo,hart2017pyomo}, which is a Python-based open-source optimization modeling tool, and MindtPy~\cite{bernal2018mixed}, which is the Mixed-Integer Nonlinear Decomposition Toolbox in Pyomo, to solve the MINLP problem. Since the optimal solution of the MINLP problem is difficult to calculate, these tools repeat the following procedure to calculate the sub-optimal solution. These tools first decomposite the MINLP problem into the continuously relaxed Non-linear Programming (NLP) problem and the Mixed-Integer Programming (MIP) problem and then calculate each problem. After calculating two problems, they consider the NLP solution as the upper bound and the MIP solution as the lower limit. They repeat the decomposition and calculation procedures until the difference between the upper and lower bounds is sufficiently small. In this paper, we use IPOPT~\cite{wachter2006implementation} and MUMPS~\cite{MUMPS:1,MUMPS:2} for solving the NLP problem and GLPK for solving the MIP problem.

As described above, the difficulty of the previous approach is that the following time-consuming tasks are required depending on the individual use cases: 
constructing a new formulation such as (18),
selecting and combining tools to solve the combined problem such as~\cite{hart2011pyomo,hart2017pyomo,bernal2018mixed,wachter2006implementation,MUMPS:1,MUMPS:2}, 
preparing the development environment, and implementing the combined problem. 
In Case \#12, the number of constructing new formulation is only one because it is the simplest use case among Cases \#1--12 and has only two control metrics.
However, when control metrics increases, the number of new formulations needing to be constructed increases by the number of control metric combinations.
On the other hand, in the proposed method, the above function can be replaced by the I/O conversion unit, which is programmable and does not need to express mathematical expressions.
In addition, it enables solutions to be calculated for various use cases.

We indicate the comparative evaluation of the performance when $N_{\rm VN} = 20$. These tools solving the MINLP problem are non-commercial and are very limited in terms of the size that can be solved. Since these tools take a long time to calculate the sub-optimal solution, this paper limits the calculation time to a maximum of 10 minutes. The proposed method uses the best CEV solutions found up to 5000 steps. As shown in Fig.~\ref{fig:time}, the calculation time of the proposed method is 10 minutes or less. Other evaluation conditions are the same as described in Section~\ref{sec:eva_conditions}.

Table~\ref{table_joint} shows the average and standard deviation ($\pm \sigma$) of the performance of the solution. We carried out 10 calculations with random initial conditions. We first set the same initial conditions for both methods. We then excluded the cases with an invalid initial solution for the proposed method and when no feasible solution is found after 10 minutes calculation for the previous method.
Results shows that our proposed method achieves better CEV compared to the previous method. Although the previous method outperforms when commercial tools are used, our proposed method almost matches up the previous method in terms of CEV, and is much easier to implement new control metrics.

\begin{table}[!t]
\renewcommand{\arraystretch}{1.3}
\caption{Performance of solution when Case \#12 and $N_{\rm VN} = 20$}
\label{table_joint}
\centering
\begin{tabular}{l|c}
\hline
{\bf Methods} & Performance (CEV) \\ 
\hline
\hline
Propose & $0.86 \pm 0.043$ \\
Previous & $0.63 \pm 0.12$ \\
\hline
\end{tabular}
\end{table}

\subsubsection{Discussion on applicability}\label{sec:applicability}
The proposed method can be applied to the VN allocation problem that considers the combination of link resource constraints (e.g., route selection) and server resource constraints (e.g., VM placement) even if the control metrics, control objective, and network model are changed as shown in Cases \#1--12. Though there are some minor constraints for the initial solution and the calculation order of each engine's evaluation, all the constraints are considered to be minor compared with the merits of the proposed method.

We first describe the condition of the initial state. As the required condition, we need to find a feasible initial solution to obtain the advantage of speeding up the solution exploration by the proposed method based on RL. Because invalid solutions are concentrated near the invalid solution, an agent of RL always obtains a negative reward and cannot learn the strategy for how to find better solutions from past exploration steps.

All evaluations in this paper assume that a feasible initial solution has already been found. When not finding a feasible solution, we try the following two ways. One is to use the proposed method against an invalid initial solution. Our proposed method can explore the feasible solution through random exploration. Fortunately, once a feasible solution is found, our method always converges to the better solution by RL. The other way is to use the sequential VN allocation, in which each VN demand is judged as to whether the physical network can allocate it or not when it is received. In this way, we allocate the $N^{\rm th}$ VM demand to the remaining resources after the $N-1$ VNs resource optimization using our proposed method. When $N^{\rm th}$ VN demand cannot be allocated to the remaining resources, this request is rejected. The evaluation of the effectiveness of the two ways is for further study.

We next describe the condition of the calculation order of each engine's evaluation values. When calculating the CEV, the evaluation values between interdependent control metrics should be calculated simultaneously (e.g., the VM placement and IDS placement), and evaluation values between dependent control metrics should be calculated sequentially (e.g., the route between VMs is determined by the VM placements). For example, in (16), the evaluation value of VM and IDS placements (i.e., $\tilde{U}^{\rm server}_{\rm max}$) is calculated by aggregating the results of VM and IDS placements. The evaluation value of route (i.e., $U^{\rm link}_{\rm max}$) is calculated after the VM and IDS placements are determined. Similarly, the evaluation value of reliability (i.e., $R^{\rm total}_t$) is calculated after the routes are determined.

\section{Conclusion}\label{conclusion}
We presented an extendable network functions virtualization (NFV)-integrated control method by coordinating multiple control algorithms. We also developed an efficient coordination algorithm on the basis of reinforcement learning (RL), which makes it possible to find better solutions with fewer explorations by learning a strategy that can improve resource-utilization efficiency with each exploration step. Simulations revealed that the proposed algorithm can improve solution exploration for $12$ representative types of the virtual network allocation use cases modeled from previous studies. This qualitatively revealed that the proposed method has extendability. We also found that it can improve resource-utilization efficiency by $22\%$ and total reliability by $8\%$ in less than $5000$ steps in the case of several hundred virtual machines (VMs) and a hundred intrusion detection systems (IDSs).

For future work, we plan to evaluate the applicability of the proposed method in more complicated use cases with realistic traffic patterns and virtual network functions (VNFs) demands. We also plan to enhance the solution-exploration-speed and scalability of our coordination algorithm by using deep RL~\cite{mnih2015human_full} and parallelization of agent learning~\cite{mnih2016asynchronous_full}.

\profile{Akito Suzuki}{received a B.E. in electronic and physical systems and an M.E. in nanoscience and engineering from Waseda University, Tokyo, Japan, in 2013 and 2015, respectively. He joined NTT Laboratories in 2015 and he has been engaged in research on network control and traffic engineering. His current research interests include network functions virtualization, mathematical optimization, and machine learning. He is a member of the IEICE.}

\profile{Ryoichi Kawahara}{received an M.E. in automatic control and a Ph.D. in telecommunication engineering from Waseda University, Tokyo, Japan, in 1992 and 2001, respectively. He joined NTT Laboratories in 1992, and has been engaged in research on traffic control for telecommunication networks, traffic measurement and analysis for IP networks, and network management, for 26 years. He is currently a professor at the Department of Information Networking for Innovation and Design, Faculty of Information Networking for Innovation and Design, Toyo University. He is a member of IEICE, IEEE, and ORSJ. He received Telecom System Technology Award from The Telecommunications Advancement Foundation in 2010, and Best Paper Awards from IEICE in 2003 and 2009.}

\profile{Masahiro Kobayashi}{is a research engineer, Communication Traffic \& Service Quality Project, NTT Network Technology Laboratories. He received his B.S. and M.S. in information science from Tohoku University, Miyagi, in 2007 and 2009. He joined NTT in 2009. He is currently studying optimal resource control in the virtualized communication network at NTT Network Technology Laboratories. He is a member of the Institute of Electronics, Information and Communication Engineers (IEICE).}

\profile{Shigeaki Harada}{received his B.S., M.S. and Ph.D. in information science from Tohoku University in 2001, 2003 and 2006, respectively. Since joining NTT Network Technology Laboratories (formerly NTT Service Integration Laboratories) in 2006, he has been engaged in research on traffic analysis and traffic control in IP networks. He is a member of IEICE.}

\profile{Yousuke Takahashi}{received his B.S. and M.S. in information science from Osaka University in 2007 and 2009. He joined NTT Laboratories in 2009 and has been engaged in researches on network management and traffic engineering. He is a member of IEICE.}

\profile{Keisuke Ishibashi}{received his B.S. and M.S. in mathematics from Tohoku University in 1993 and 1995, respectively, and received his Ph.D. in information science and technology from the University of Tokyo in 2005.  From 1995 to 2018, he worked at NTT Laboratories where he was involved in the research on the measurement and analysis of internet traffic and performance. He is currently an associate professor of information science at the International Christian University. He is a member of the IEICE, IEEE and Japan Society for Software Science and Technology. He received  Best Paper Awards from IEICE in 2019.}

\end{document}